\documentclass[aps,prb,floatfix,twocolumn,superscriptaddress]{revtex4-2}

\usepackage{mathrsfs}
\usepackage{graphicx}
\usepackage[english]{babel}
\usepackage{amsmath}
\usepackage{amssymb}
\usepackage{xcolor}
\usepackage{cancel}
\usepackage[normalem]{ulem}

\usepackage{relsize}
\usepackage{CJK}
\usepackage{dsfont}
\usepackage{bm}

\newcommand{\me}{\mathrm{e}}
\newcommand{\mi}{\mathrm{i}}

\newcommand{\dif}{\mathrm{d}}

\allowdisplaybreaks

\begin{document}

\title{Uhlmann and scalar Wilczek-Zee phases of degenerate quantum systems}

\author{Xin Wang}
\affiliation{School of Physics, Southeast University, Jiulonghu Campus, Nanjing 211189, China}

\author{Xu-Yang Hou}
\affiliation{School of Physics, Southeast University, Jiulonghu Campus, Nanjing 211189, China}

\author{Hao Guo}
\email{guohao.ph@seu.edu.cn}
\affiliation{School of Physics, Southeast University, Jiulonghu Campus, Nanjing 211189, China}
\affiliation{Hefei National Laboratory, Hefei 230088, China}

\author{Chih-Chun Chien}
\email{cchien5@ucmerced.edu}
\affiliation{Department of physics, University of California, Merced, CA 95343, USA}

\begin{abstract}
The Wilczek-Zee (WZ) holonomy arises in degenerate states while the Uhlmann holonomy characterizes finite-temperature topology. We investigate possible relationships between the Uhlmann phase and the scalar WZ phase, which reflects the Uhlmann and WZ holonomy respectively, in an exemplary four-level model with two doubly degenerate subspaces. Through exact solutions, we contrast the behavior of the Uhlmann and WZ connections and their associated phases. In the zero-temperature limit, the Uhlmann phase may or may not agree with the scalar WZ phase of the degenerate ground states due to obstructions from the Hamiltonian manifested as Dirac points. This is in stark contrast to non-degenerate systems where the correspondence between the Uhlmann and Berry phases in general holds. Our analyses further show that for the example studied here, the Uhlmann phase catches the singular behavior at the Dirac points while the WZ connection and scalar WZ phase vanish along a zero-field axis. We also briefly discuss possible experimental implications.
\end{abstract}

\maketitle

\section{Introduction}
Geometric phases serve as a cornerstone in the study of interesting topological properties of quantum systems. These phases arise from the evolution of a quantum state in the parameter space and reflect the underlying geometric structures, depending only on the path taken rather than the dynamic details of the evolution. The original idea was introduced via the Berry phase \cite{Berry84}, which characterizes the phase accumulated by a non-degenerate quantum state during an adiabatic and cyclic evolution. Since its introduction, the Berry phase has become a stepping stone for understanding a wide range of topological phenomena, ranging from the quantum Hall effect to topological insulators~\cite{TKNN,Haldane,KaneRMP,ZhangSCRMP,MooreN,MoorePRB,FuLPRL,Bernevigbook,ChiuRMP,KaneMele,KaneMele2,BernevigPRL}.

For quantum systems with degeneracy, the geometric phase exhibits even richer physics. The Wilczek-Zee (WZ) phase of degenerate systems~\cite{PhysRevLett.52.2111} extends the Berry phase of non-degenerate systems. Unlike the scalar Berry phase, the WZ phase manifests as a unitary matrix acting on the degenerate subspace, reflecting the underlying non-Abelian holonomy. This non-Abelian nature introduces additional complexity and enables the study of topological properties in degenerate quantum systems \cite{Anandan_1988,Cheon_2009,PhysRevA.71.012110,PhysRevA.93.012116}. Moreover, the WZ phase has found applications across diverse areas, including nuclear rotations of diatomic molecules \cite{PhysRevLett.56.893,PhysRevLett.56.2779}, nuclear quadrupole resonance (NQR) \cite{PhysRevLett.60.2734,PhysRevA.42.3107}, dynamics of deformable bodies \cite{Shapere_1989,shapere1989gauge}, molecular Kramers doublets \cite{PhysRevLett.59.161}, semiconductor heterostructures \cite{PhysRevB.57.12302}, ion traps \cite{PhysRevA.59.2910}, etc. More recently, it has also been experimentally observed using atomic Bose-Einstein condensates \cite{Sugawa_2021}.

While pure states described by wavefunctions are well-suited for studying geometric phases of equilibrium systems at zero temperature, finite-temperature systems are ubiquitous with thermal effects giving rise to mixed states described by density matrices. It is thus crucial to address how geometric phases can be generalized to mixed states. The Uhlmann phase \cite{Uhlmann86, Uhlmann91} provides a rigorous solution by extending the concept of geometric phase to density matrices through the definition of the parallel-transport condition of purification of density matrices. The Uhlmann phase has been instrumental in revealing topological phase transitions at finite temperatures in various systems \cite{Viyuela14, ViyuelaPRL14-2, PhysRevLett.113.076407, OurPRA21, Morachis_Galindo_2021, OurPRB20b, Zhang21,PhysRevB.110.035144,PhysRevB.97.235141,OurUhlmannQuench}. Despite its utility, the Uhlmann phase remains less understood than its pure-state counterparts, particularly when it comes to its relationship in the zero-temperature limit to the Berry and WZ phases.

Previous research has revealed interesting examples where the Uhlmann phase reduces, in the zero-temperature limit, to the Berry phase at zero temperature in non-degenerate systems \cite{Viyuela14, Morachis_Galindo_2021, OurUB}, despite their different origins from topologically distinct principal fiber bundles.
By presenting a concrete example that exhibits a violation of the correspondence for degenerate cases, we will identify topological obstructions due to the presence of Dirac points when the Hamiltonian vanishes and the system becomes completely degenerate.
This dichotomy motivates a broader examination of how finite-temperature geometric phases relate to possible zero-temperature counterparts, especially in the presence of degeneracy. However, it may not be meaningful to directly compare the Uhlmann phase with the matrix-valued WZ phase in the presence of degeneracy. To facilitate a fair comparison with the geometric phase from the WZ holonomy of degenerate ground states, we propose to use the scalar Wilczek-Zee phase, which has a similar definition as the Uhlmann phase to reflect the underlying non-Abelian holonomy, instead of the matrix-valued WZ phases in the search of possible relations with the Uhlmann phase in the zero-temperature limit.

To provide a concrete analysis, we study a four-level model with two doubly degenerate subspaces, which readily exhibits a variety of phenomena for contrasting the behavior of the Uhlmann and scalar WZ phases. Moreover, we explicitly derive the Uhlmann connection and the WZ connection that generate the corresponding non-Abelian holonomy. Through two simple examples of the four-level model, we demonstrate some agreements and disagreements by varying the underlying parameter spaces between the Uhlmann phase in the zero-temperature limit and the scalar WZ phase. An attempt to analytically understand the zero-temperature limit of the Uhlmann holonomy and phase in contrast to the WZ holonomy and its scalar phase further elucidates some interesting but challenging topological properties behind degenerate quantum systems at and beyond zero temperature, particularly in the presence of singular behavior from the Hamiltonian.

The rest of the paper is organized as follows. Section~\ref{Sec2} introduces the four-level model with two doubly degenerate subspaces and its properties, defines the scalar WZ phase, and summarizes the explicit expressions for the Uhlmann and WZ connections along with their associated holonomy and phases. Section~\ref{Sec3} examines two specific examples of the four-level model to demonstrate possible relationships between the Uhlmann phase in the zero-temperature limit and the scalar WZ phase. Section~\ref{Sec4} provides a theoretical analysis of the Uhlmann phase in the zero-temperature limit and proposes possible conditions for reduction to the scalar WZ phase in degenerate systems. Section~\ref{Sec:PhaseReduction} analyzes the existence of a zero-field axis where the WZ connection vanishes and the Uhlmann connection near a Dirac point in the 4D tight-binding model to pinpoint the difference between the zero-temperature limit of the Uhlmann phase and the scalar WZ phase. Section~\ref{Sec5} discusses some implications in possible experiments and simulations. Finally, Section~\ref{Sec6} concludes our work. The Appendix summarizes some 
details and subtleties in the definitions and evaluations of 
various geometric phases, presents a geometric proof of the zero-field axis condition, along with a conditional proof of the Uhlmann-Berry correspondence for non-degenerate systems.

\section{Four-level model with two doubly degenerate spaces}\label{Sec2}
\subsection{The Hamiltonian}
We consider a generic four-level model with two doubly degenerate spaces described by the Hamiltonian
\begin{align}\label{H}
H=\sum_{i=1}^5R_i\Gamma^i.
\end{align}
Here the gamma matrices are defined by $\Gamma^i=\sigma_1\otimes\sigma_i$ for $i=1,2,3$, $\Gamma^4=\sigma_2\otimes 1_2$ and $\Gamma^5=\sigma_3\otimes 1_2$, where $\sigma_i$ with $i=1,2,3$ denote the Pauli matrices and $1_2$ denotes the $2\times 2$ identity matrix. These gamma matrices satisfy the Clifford algebra $\{\Gamma^i,\Gamma^j\}=2\delta_{ij}1_4$
with $1_4$ being the $4\times 4$ identity matrix. Throughout the paper, we set $\hbar=1=k_B$.

The four eigenstates form two pairs of degenerate states and are given by
\begin{align}\label{EL1}
|\psi_{a,c}\rangle&=\frac{1}{\sqrt{2R(R\mp R_5)}}\begin{pmatrix}
-R_3+\mi R_4\\
-R_1-\mi R_2\\
R_5\mp R\\
0
\end{pmatrix},
\notag\\
|\psi_{b,d}\rangle&=\frac{1}{\sqrt{2R(R\mp R_5)}}\begin{pmatrix}
-R_1+\mi R_2\\
R_3+\mi R_4\\
0\\
R_5\mp R
\end{pmatrix}
\end{align} with $R=\sqrt{\sum_{i=1}^5R^2_i}$.
They satisfy the eigenvalue equations \begin{align}\label{HE} H|\psi_{a,b}\rangle &= +R|\psi_{a,b}\rangle, \quad H|\psi_{c,d}\rangle = -R|\psi_{c,d}\rangle, \end{align}
where each level is twofold degenerate. We introduce the projectors onto the subspaces with eigenvalues $\pm R$ as
\begin{align}\label{P1}
P_+ &= |\psi_a\rangle\langle\psi_a| + |\psi_b\rangle\langle\psi_b| = \frac{1}{2}\left(1_4 + \hat{R}_i \Gamma^i\right), \notag\\
P_- &= |\psi_c\rangle\langle\psi_c| + |\psi_d\rangle\langle\psi_d| = \frac{1}{2}\left(1_4 - \hat{R}_i \Gamma^i\right),
\end{align}
where $\hat{R}_i = R_i / R$.
Note that each expression for $P_\pm$ corresponds to a different representation: the first form is written in the eigen-basis of the Hamiltonian, while the second form is given in terms of the Gamma matrices.  In thermal equilibrium, the density matrix is
\begin{align}\label{rho1}
\rho=&\frac{\me^{-\beta R}}{Z}P_++\frac{\me^{\beta R}}{Z}P_-=\frac{1}{4}\left(1_4-\tanh(\beta R)\hat{R}_i\Gamma^i\right),
\end{align}
where $\beta=\frac{1}{T}$ is the inverse temperature, and $Z=2\me^{-\beta R}+2\me^{\beta R}=4\cosh(\beta R)$ is the partition function.

\subsection{Uhlmann holonomy}
In general, we consider a quantum system with a $N$-dimensional Hilbert space. A statistical ensemble is represented by a mixed state, which is described by a density matrix. To define the geometric phase of mixed states, one can introduce the notion of the amplitude through purification of a full-rank density matrix via $W=\sqrt{\rho}\mathcal{U}$, where $\mathcal{U}\in U(N)$ is a unitary phase factor. Conversely, the density matrix can be reconstructed as $\rho=WW^\dag$, which is independent of the choice of the phase factor. For a system in thermal equilibrium described by the canonical ensemble, the density matrix has the diagonal form $\rho=\sum_n\lambda_n|n\rangle\langle n|$ in the eigen-basis of the Hamiltonian. The purification of the density matrix is then expressed as $W=\sum_n\sqrt{\lambda_n}|n\rangle\langle n|\mathcal{U}$. One can introduce a pure-state like representation of $W$, called the purified state, as $|W\rangle=\sum_n\sqrt{\lambda_n}|n\rangle \otimes \mathcal{U}^T|n\rangle $. In this way, the Hilbert-Schmidt product introduces an inner product between purified states: $\langle W_1|W_2\rangle=\text{Tr}(W^\dag_1W_2)$ \cite{Bengtsson_book}.

For the four-level model described by Eq.~\eqref{H}, we define $\mathbf{R} = (R_1,R_2,R_3,R_4,R_5)^{T}$ and consider the system undergoing cyclic evolution along a closed path $C$ described by $\mathbf{R}(t)$ with $0\le t\le \tau$ and $\mathbf{R}(0)=\mathbf{R}(\tau)$ in the parameter space. For simplicity, we denote $\rho(t)\equiv \rho(\mathbf{R}(t))$. If the amplitude $W(t)$ satisfies the Uhlmann parallel-transport condition \begin{align}\label{pc}
W^\dag\dot{W}=\dot{W}^\dag W,
\end{align}
then $W(t)$ is called a horizontal lift of $\rho(t)$.
The condition further implies the differential equation
 \begin{align}\label{U}
\dot{\mathcal{U}}=-\mathcal{A}_\text{U}\mathcal{U},
\end{align}
where $\mathcal{A}_\text{U}$ is the Uhlmann connection.
In general, the curve $W(t)$ is not closed, as the initial and final phase factors may differ by the Uhlmann holonomy, which is obtained from Eq.~(\ref{U}) and has the form
 \begin{align}\label{Eq:Uholonomy}
\mathcal{U}(\tau)=\mathcal{P}\me^{-\mathlarger{\oint}_C\mathcal{A}_\text{U}}\mathcal{U}(0),
\end{align}
where $\mathcal{P}$ denotes path ordering. Using $\rho=\sum_i\lambda_i|i\rangle\langle i|$, $\mathcal{A}_\text{U}$ can be expressed as
\begin{align}\label{AUE}
\mathcal{A}_\text{U}=-\sum_{ij}|i\rangle\frac{\langle i|[\mathrm{d}\sqrt{\rho},\sqrt{\rho}]|j\rangle}{\lambda_i+\lambda_j}\langle j|.
\end{align}
For four-level model featuring two doubly degenerate subspaces introduced in Eq.~\eqref{H}, a direct evaluation of Eq.~(\ref{AUE}) shows that the Uhlmann connection simplifies to
\begin{align}\label{M1}
&\mathcal{A}_{\text{U}} =-\left[1-\text{sech}\left(\frac{R}{T}\right)\right]\left(P_+\dif P_-+P_-\dif P_+\right)\notag\\
=&\frac{1-\text{sech}\left(\frac{R}{T}\right)}{2}\hat{R}_a\dif\hat{R}_b\Gamma^a\Gamma^b\notag\\
=&-\frac{\mi}{4}\left[1-\text{sech}\left(\frac{R}{T}\right)\right](\hat{R}_a\dif\hat{R}_b-\hat{R}_b\dif\hat{R}_a)\Gamma^{ab},
\end{align}
where $\Gamma^{ab}=\frac{\mi}{2}[\Gamma^a,\Gamma^b]$. More details are given in Appendix \ref{app1}. One important distinction between Eqs.~(\ref{AUE}) and (\ref{M1}) is that Eq.~(\ref{AUE}) is given in the eigen-basis of the Hamiltonian, which is also the eigen-basis of the canonical-ensemble density matrix in equilibrium. In contrast, Eq.~(\ref{M1}) is written in the Gamma-matrix representation with known forms. The latter allows for a direct calculation of the Uhlmann connection in terms of the Gamma matrices:
\begin{align}\label{AUM}
\mathcal{A}_{\text{U}} =-\frac{1-\text{sech}\left(\frac{R}{T}\right)}{2R^2}M,\end{align}
where the matrices elements of $M$ are given by
\begin{small}
\begin{align}\label{M}
M_{11}&=\mi (R_2 \dif R_1 - R_1 \dif R_2 + R_4 \dif R_3 - R_3 \dif R_4),\notag\\
M_{12}&=(R_1 - \mi R_2) (\dif R_3 - \mi \dif R_4) - (R_3 - \mi R_4) (\dif R_1 - \mi \dif R_2),\notag\\
M_{13}&=-R_5 \dif R_3 + \mi R_5 \dif R_4 + (R_3 - \mi R_4) \dif R_5,\notag\\
M_{14}&=-R_5 \dif R_1 + \mi R_5 \dif R_2 + (R_1 - \mi R_2) \dif R_5,\notag\\
M_{21}&=(R_3 + \mi R_4) (\dif R_1 + \mi \dif R_2) - (R_1 + \mi R_2) (\dif R_3 + \mi \dif R_4),\notag\\
M_{22}&=\mi (-R_2 \dif R_1 + R_1 \dif R_2 - R_4 \dif R_3 + R_3 \dif R_4),\notag\\
M_{23}&=(R_1 + \mi R_2) \dif R_5 - R_5 (\dif R_1 + \mi \dif R_2),\notag\\
M_{24}&=R_5 \dif R_3 + \mi R_5 \dif R_4 - (R_3 + \mi R_4) \dif R_5,\notag\\
M_{31}&=R_5 \dif R_3 + \mi R_5 \dif R_4 - (R_3 + \mi R_4) \dif R_5,\notag\\
M_{32}&=R_5 \dif R_1 - \mi R_5 \dif R_2 - (R_1 - \mi R_2) \dif R_5,\notag\\
M_{33}&=\mi (R_2 \dif R_1 - R_1 \dif R_2 - R_4 \dif R_3 + R_3 \dif R_4) ,\notag\\
M_{34}&=(R_1 - \mi R_2) (\dif R_3 + \mi \dif R_4) - (R_3 + \mi R_4) (\dif R_1 - \mi \dif R_2),\notag\\
M_{41}&=R_5 \dif R_1 + \mi R_5 \dif R_2 - (R_1 + \mi R_2) \dif R_5 ,\notag\\
M_{42}&= -R_5 \dif R_3 + \mi R_5 \dif R_4 + (R_3 - \mi R_4) \dif R_5,\notag\\
M_{43}&= (R_3 - \mi R_4) (\dif R_1 + \mi \dif R_2) - (R_1 + \mi R_2) (\dif R_3 - \mi \dif R_4),\notag\\
M_{44}&=\mi (-R_2 \dif R_1 + R_1 \dif R_2 + R_4 \dif R_3 - R_3 \dif R_4).
\end{align}
\end{small}
At the end of the parallel transport along the loop $C$ in the parameter space, the mixed state acquires the Uhlmann phase
\begin{align}\label{thetaU}
\theta_{\text{U}}(C)=\arg\langle W(0)|W(\tau)\rangle=\arg\text{Tr}\left[\rho(0)\mathcal{P}\me^{-\mathlarger{\oint}_C \mathcal{A}_{\text{U}}}\right].
\end{align}
When $T\rightarrow 0$, a system in thermal equilibrium should approach its ground state. Later on, we will analyze the behavior of the Uhlmann connection, Uhlmann holonomy, and Uhlmann phase in the zero-temperature limit in the presence of a degenerate ground-state subspace.

\subsection{Wilczek-Zee holonomy}
At zero temperature, the ground states are of interest and in equilibrium, they are pure states. Here we briefly summarize the pure-state holonomy in the presence of degenerate states by  considering a system evolving adiabatically along a closed path $C$ parameterized by $\mathbf{R}(t)$ for $t \in [0, \tau]$ with $\mathbf{R}(0) = \mathbf{R}(\tau)$. Consequently, the Hamiltonian $H(t) \equiv H(\mathbf{R}(t))$ returns to its initial form at $t = \tau$. Under the cyclic evolution, the pure states in each degenerate subspace undergo a unitary transformation:
\begin{align}\label{WZc}
|\psi_i(\tau)\rangle &= \sum_{j=a,b} \bigl(U_{+}\bigr)_{ij}|\psi_j(0)\rangle,
\quad i=a,b,\notag\\
|\psi_i(\tau)\rangle &= \sum_{j=c,d} \bigl(U_{-}\bigr)_{ij}|\psi_j(0)\rangle,
\quad i=c,d.
\end{align}
In the ordered basis $\{|\psi_a\rangle,|\psi_b\rangle,|\psi_c\rangle,|\psi_d\rangle\}$, the full holonomy, which gives the Wilczek-Zee matrix \cite{PhysRevLett.52.2111}, factorizes into a block-diagonal form:
\begin{align}\label{UWZ}
U =
\begin{pmatrix}
U_{-} & 0 \\
0 & U_{+}
\end{pmatrix}=
\begin{pmatrix}
\mathcal{P}\me^{\mi\mathlarger{\int}_{0}^{\tau}\mathbf{A}_{-}\cdot\dot{\mathbf{R}}\dif t} & 0 \\
0 & \mathcal{P}\me^{\mi\mathlarger{\int}_{0}^{\tau}\mathbf{A}_{+}\!\cdot\dot{\mathbf{R}}\dif t}
\end{pmatrix},
\end{align}
where the elements of the WZ potentials are given by
$\mathbf{A}_{+ij}=\mi\langle\psi_i|\nabla_\mathbf{R}|\psi_j\rangle$ for $i,j=a,b$ and $\mathbf{A}_{-ij}=\mi\langle\psi_i|\nabla_\mathbf{R}|\psi_j\rangle$ for $i,j=c,d$.

In order to directly compare with the Uhlmann connection later, we define the WZ connection as
\begin{align}
\mathcal{A}_{\pm,ij}=\langle\psi_i|\dif|\psi_j\rangle=-\mi\mathbf{A}_{\pm,ij}\cdot\dif \mathbf{R}
 \end{align}
 for $i$, $j$=$a$, $b$ or $c$, $d$. This definition differs from the conventional one by a factor of $\mi=\sqrt{-1}$, but it does not change any physical essence. Thus, the total WZ connection matrix is \begin{align}\label{AWZ}\mathcal{A}_\text{WZ}=\begin{pmatrix}\mathcal{A}_- & 0 \\ 0 & \mathcal{A}_+\end{pmatrix},\end{align} and the corresponding WZ holonomy along a loop $C$ is given by
$U(C) =\mathcal{P}\me^{-\mathlarger{\oint}_C\mathcal{A}_\text{WZ}}$.
Unlike the Uhlmann connection given in Eq.~(\ref{M1}), here the WZ connection is, by definition, expressed in the eigen-basis of the Hamiltonian. To express the WZ connection by the Gamma matrices typically requires a more involved calculation. Nevertheless, we obtain
\begin{align}\label{A+-ac12}
\mathcal{A}_{+aa,-cc}&=-\mathcal{A}_{+bb,-dd}\notag\\
&=\dfrac{R_2\dif R_1-R_1\dif R_2-R_4\dif R_3+R_3\dif R_4}{2\mi R(R\mp R_5)}
\end{align}
and
\begin{widetext}
 \begin{align}\label{A+-ab12}
\mathcal{A}_{+ab,-cd}&=\mathcal{A}^*_{+ba,-dc}=\frac{R_1\dif R_4-R_4\dif R_1-R_2\dif R_3+R_3\dif R_2+\mi(R_1\dif R_3-R_3\dif R_1+R_2\dif R_4-R_4\dif R_2)}{2\mi R(R\mp R_5)}.
\end{align}
\end{widetext}

\subsection{Scalar Wilczek-Zee phase}
Since the explicit forms of $\mathcal{A}_\text{U}$ and $\mathcal{A}_\text{WZ}$ are available, it is intriguing to check the $T\to0$ limit of $\mathcal{A}_\text{U}$ and see if there is any relationship with $\mathcal{A}_\text{WZ}$. A key distinction, however, is that they have been derived in different bases: Eq.~(\ref{M1}) presents the Uhlmann connection in the Gamma-matrix representation, whereas Eq.~(\ref{UWZ}) gives the WZ connection in the eigen-basis of the Hamiltonian. This apparent discrepancy seems to preclude a direct comparison.
Nevertheless, we demonstrate in Appendix~\ref{app1} that the Uhlmann connection can also be rewritten in terms of the eigen-basis of the Hamiltonian.
Consequently, Eq.~(\ref{AUE}) can be rewritten as \cite{OurUB}
\begin{align}\label{AUE2}
\mathcal{A}_\mathrm{U}
&=-\sum_{i,j}\frac{\bigl(\sqrt{\lambda_i}-\sqrt{\lambda_j}\bigr)^2}{\lambda_i+\lambda_j}
|i\rangle\langle i|\dif|j\rangle\langle j|
\end{align}
in the eigen-basis of the Hamiltonian.
Whenever $|i\rangle$ and $|j\rangle$ lie in the same degenerate subspace, $\lambda_i=\lambda_j$, causing each diagonal block of $\mathcal{A}_\mathrm{U}$ to vanish identically.  All nonzero entries of $\mathcal{A}_\mathrm{U}$ therefore arise from the coupling between different energy levels. By contrast, $\mathcal{A}_\mathrm{WZ}$ is explicitly block-diagonal, with its only nonzero elements confined within each degenerate subspace.

Since an explicit relation between $\mathcal{A}_\text{U}$ in the $T\rightarrow 0$ limit and $\mathcal{A}_\text{WZ}$ is difficult to obtain in general, we shift our focus to the Uhlmann and WZ phases arising from the corresponding holonomy. However, the conventional WZ phase is matrix-valued~\cite{PhysRevLett.52.2111}, as shown by the elements in Eq.~(\ref{UWZ}). To enable a direct comparison with the zero-temperature limit of the Uhlmann phase, we introduce the scalar Wilczek-Zee phase associated with the degenerate ground-state subspace:
\begin{align}\label{thetaWZ}
\theta_{\text{WZ}}(C) = \arg \text{Tr}_- \left( \frac{1}{2} P_- (0)U_-(C) \right),
\end{align}
where $\text{Tr}_-$ denotes the trace over the degenerate ground-state subspace, and the prefactor $\frac{1}{2}$ ensures the normalization $\text{Tr}_- \left( \frac{1}{2} P_- \right) = 1$. Although the scalar WZ phase can be similarly defined for the degenerate excited-state subspace, it is irrelevant to the zero-temperature limit of the Uhlmann phase. Therefore, we will focus on the scalar WZ phase of the degenerate ground states in the following.
In the absence of degenerate ground states ($D = 1$), $\theta_{\text{WZ}}$ reduces to the Berry phase, as expected. When the ground state is $D$-fold degenerate, the normalization factor $\frac{1}{2}$ should be generalized to $\frac{1}{D}$.

We remark that the Uhlmann and scalar WZ phases represent the phase of the average of the holonomy over the density matrix and degenerate states, respectively. The two scalar phases facilitate a direct comparison despite the different underlying holonomy structures. We caution that when a zero scalar phase arises from a nontrivial holonomy, it implies the averaged holonomy leads to a positive real number and thus does not contribute to the phase.

To make the comparison between the $T\rightarrow 0$ Uhlmann phase and scalar WZ phase more transparent, we define the analogue of purification of the ground-state projector as $W_- = \frac{1}{\sqrt{D}} \sqrt{P_-} \mathcal{U}$, where $\mathcal{U}$ represents the phase factor associated with the degenerate ground-state subspace of dimension $D$.
Under cyclic evolution along a closed path $C(t)$ ($0 \leq t \leq \tau$) in the parameter space, the initial and final phase factors differ by a WZ holonomy:
 \begin{align}\label{Eq:WZholonomy}
\mathcal{U}(\tau)=\mathcal{P}\me^{-\mathlarger{\oint}_C\mathcal{A}_-}\mathcal{U}(0)=U_-(C)\mathcal{U}(0),
\end{align}
which is the counterpart of the Uhlmann holonomy in Eq.~(\ref{Eq:Uholonomy}). In a direct analogy to the Uhlmann phase defined in Eq.~(\ref{thetaU}), we express the scalar WZ phase as:
\begin{align}\label{Eq:sWZ}
\theta_{\text{WZ}}(C) = \arg  \langle W_-(0) | W_-(\tau) \rangle,
\end{align}
which exactly coincides with Eq.~(\ref{thetaWZ}) (see Appendix \ref{app2a} for the consistency of the definitions).
Here the analogies of the purified state $| W_-(t) \rangle$ and the inner products are also introduced in a similar fashion. Since the density matrix approaches $\rho \to \frac{1}{D} P_-$ as $T \to 0$, the formally similar expressions of the Uhlmann and scalar WZ phases warrant some investigations into their relations. We set out to search for any correspondence between the Uhlmann phase in the zero-temperature limit and the scalar WZ phase in the four-level model described by Eq.~\eqref{H} because such a correspondence may serve as a bridge to link the topological properties of mixed states with those of degenerate ground states.

\section{Examples}\label{Sec3}
Here we present two examples of the four-level model discussed in Sec.~\ref{Sec2} to contrast the $T\rightarrow 0$ limit of the Uhlmann phase and the WZ phase of the degenerate ground states.

\subsection{A simple illustration}
We begin by considering a simple case with $R_1 = R_3 = \frac{R}{\sqrt{2}} \sin\theta \cos\phi$, $R_2 = R_4 = \frac{R}{\sqrt{2}} \sin\theta \sin\phi$, and $R_5 = R \cos\theta$ in Eq.~(\ref{H}).
Consequently, the parameter manifold is a two-dimensional spherical surface parameterized by $(\theta, \phi)$.
After substituting the parameters into Eq.~(\ref{M}) and following a straightforward calculation, the Uhlmann connection is given by
\begin{align}
\mathcal{A}_{\text{U}} =&- \frac{1-\operatorname{sech}\left(\frac{R}{T}\right) }{2}\notag\\ \times&\begin{pmatrix}
-\mi \sin^2\theta \dif\phi & 0 & -\frac{\me^{-\mi\phi}}{\sqrt{2}}  \Omega & -\frac{\me^{-\mi\phi}}{\sqrt{2}}  \Omega \\
0 & \mi \sin^2\theta \dif\phi & -\frac{\me^{\mi\phi}}{\sqrt{2}}  \Omega^* & \frac{\me^{\mi\phi}}{\sqrt{2}}  \Omega^* \\
\frac{\me^{\mi\phi}}{\sqrt{2}}  \Omega^* & \frac{\me^{-\mi\phi}}{\sqrt{2}}  \Omega & 0 & \mi \sin^2\theta \dif\phi \\
\frac{\me^{\mi\phi}}{\sqrt{2}}  \Omega^* & -\frac{ \me^{-\mi\phi}}{\sqrt{2}} \Omega & \mi \sin^2\theta \dif\phi & 0
\end{pmatrix},
\end{align}
where $\Omega = \dif\theta - \mi \cos\theta \sin\theta \dif\phi$ is a 1-form, and $\Omega^*$ is its complex conjugate.
For convenience, we consider a closed path taken along the equator (i.e., $\theta = \frac{\pi}{2}$) of the parameter space, starting from the point $(\theta(0),\phi(0))=(\frac{\pi}{2},0)$. Along the path, the Uhlmann connection simplifies to
\begin{align}\label{AUtoy}
\mathcal{A}_{\text{U}}&=\mi\frac{1-\operatorname{sech}\left(\frac{R}{T}\right) }{2}\begin{pmatrix}
1 & 0 & 0 & 0 \\
0 & -1 & 0 & 0 \\
0 & 0 & 0 & -1 \\
0 & 0 & -1 & 0
\end{pmatrix}\dif \phi\notag\\
&=\mi\frac{1-\operatorname{sech}\left(\frac{R}{T}\right) }{2}\sigma_3 \oplus (-\sigma_1)\dif \phi,
\end{align}
which is proportional to a constant matrix.

As a consequence, the path-ordering operator $\mathcal{P}$ in Eq.~(\ref{thetaU}) becomes straightforward and may be omitted in the evaluation of the Uhlmann phase. Furthermore, the absence of the off-diagonal blocks of $\mathcal{A}_{\text{U}}$ implies that under this configuration, the contributions to the connection from the coupling between the ground and excited states vanish. To evaluate the Uhlmann phase, we also notice that the initial density matrix is given by
\begin{align}\label{rho0}
	&\rho(0)=\frac{1}{Z}\me^{-\beta H(\theta(0),\phi(0))}\notag\\
=&\left(
	\begin{array}{cccc}
		\frac{1}{4} & 0 & -\frac{\tanh (\beta  R)}{4 \sqrt{2}} & -\frac{\tanh (\beta  R)}{4 \sqrt{2}} \\
		0 & \frac{1}{4} & -\frac{\tanh (\beta  R)}{4 \sqrt{2}} & \frac{\tanh (\beta  R)}{4 \sqrt{2}} \\
		-\frac{\tanh (\beta  R)}{4 \sqrt{2}} & -\frac{\tanh (\beta  R)}{4 \sqrt{2}} & \frac{1}{4} & 0 \\
		-\frac{\tanh (\beta  R)}{4 \sqrt{2}} & \frac{\tanh (\beta  R)}{4 \sqrt{2}} & 0 & \frac{1}{4} \\
	\end{array}
	\right).
\end{align}
Introducing $\chi \equiv 1 - \operatorname{sech}(R/T)$, the Uhlmann holonomy is evaluated as
\begin{align}
	\mathcal{P}\me^{-\oint_C\mathcal{A}_\text{U}}
		&= \mathcal{P} \exp\left(-\mi \oint_C \frac{\chi}{2}  \dif \phi \sigma_3 \oplus( -\sigma_1)\right) \notag \\
		&= \exp(-\mi \pi \chi \sigma_3) \oplus \exp(\mi \pi \chi \sigma_1) \notag \\
		&= \begin{pmatrix}
			\me^{-\mi \pi \chi} & 0 & 0 & 0 \\
			0 & \me^{\mi \pi \chi} & 0 & 0 \\
			0 & 0 & \cos(\pi \chi) & \mi \sin(\pi \chi) \\
			0 & 0 & \mi \sin(\pi \chi) & \cos(\pi \chi)
		\end{pmatrix},
\end{align}
Using Eq.~(\ref{rho0}), the corresponding Uhlmann phase is then given by
	\begin{equation}\label{thetaU2}
		\theta_{\text{U}}(C) = \arg\text{Tr}\left[\rho(0) 	\mathcal{P}\me^{-\oint_C\mathcal{A}_\text{U}}\right] = \arg\left(\cos(\pi \chi)\right).
	\end{equation}

For the WZ connection in the degenerate ground-state subspace, it can be shown that $\mathcal{A}_\pm=\frac{1\pm\cos\theta}{2 \mi}\dif\phi\sigma_1$ for this simple case. If the same path along the equator is followed with $\theta=\frac{\pi}{2}$, the WZ connection of the degenerate ground states reduces to
\begin{align}\label{A-}
\mathcal{A}_-=-\frac{\mi}{2 }\begin{pmatrix}0 & 1\\ 1& 0\end{pmatrix}\dif\phi.\end{align}
Again, the path-ordering can be omitted since the connection is proportional to a constant matrix. The corresponding holonomy of the degenerate ground states is given by
\begin{align}
U_-=\me^{\frac{\mi}{2}\sigma_1\mathlarger{\oint}\dif\phi}=-1_2.
\end{align}
Using Eq.~(\ref{thetaWZ}), we find that
\begin{align}
\theta_{\text{WZ}}=\arg\text{Tr}_-\left(-\frac{1}{2}P_-\right)=\arg(-1)=\pi.
\end{align}
As a comparison, when $T\to 0$, we have $\beta\to \infty$ and $\chi\to 1$. Thus, Eq.~(\ref{thetaU2}) implies
\begin{align}
	\lim_{T\to 0}\theta_{\text{U}}=\arg  (\cos(\pi)) =\pi=\theta_{\text{WZ}}.
\end{align}
Therefore, this simple case shows a correspondence between the $T\rightarrow 0$ Uhlmann phase and the scalar WZ phase.

\begin{figure}[t]
	\centering
	\includegraphics[width=3.2in,clip]{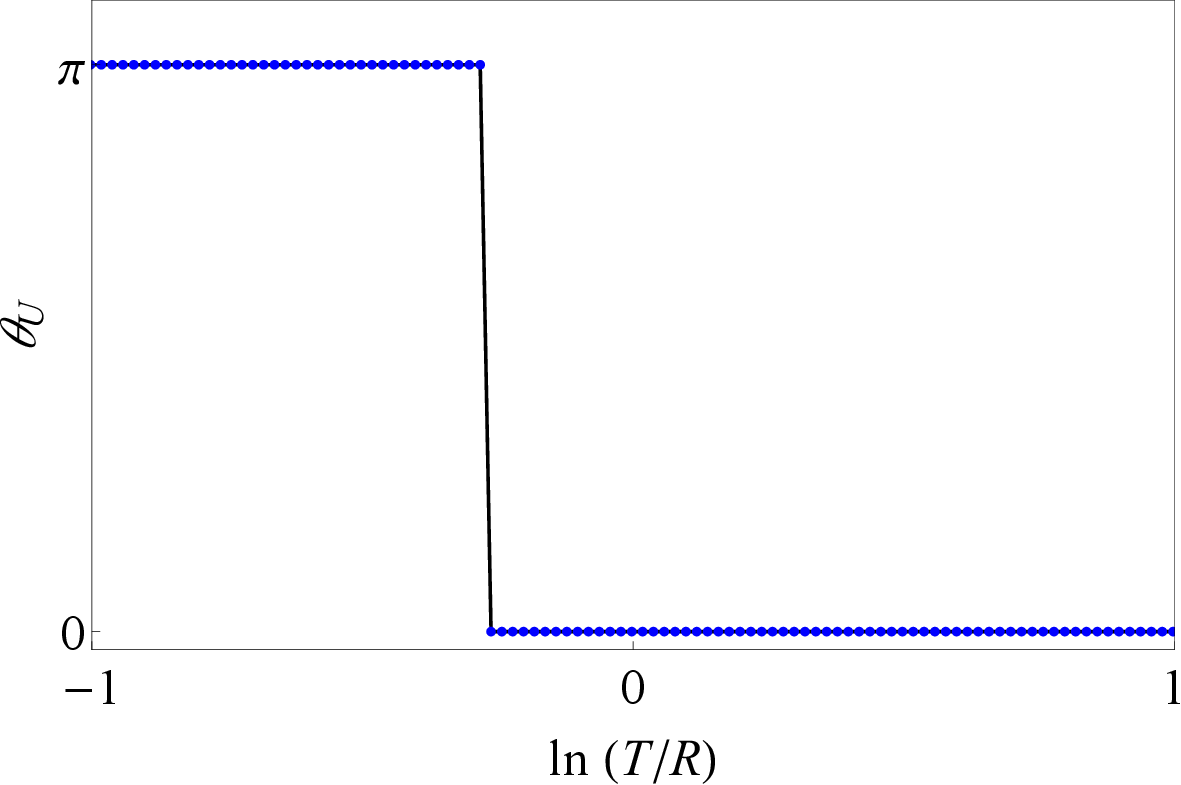}\\
	\caption{Uhlmann phase 	$\theta_{\text{U}}$ as a function of temperature $T$ for the simple case. A $\pi$-jump occurs at $T\approx0.75R$.}
	\label{Fig1}
\end{figure}

Before proceeding further, we note that the Uhlmann phase can signal finite-temperature topological phase transitions~\cite{Viyuela14, OurPRA21, Morachis_Galindo_2021} with quantized jumps. At infinitely high temperatures, the density matrix becomes proportional to the identity matrix, implying $\rho \propto 1_4$ for a four-level system. As a result, any evolution of $\rho(t)$ along a loop in the parameter space effectively collapses to a single point, thereby rendering the Uhlmann phase trivial~\cite{OurPRA21}. Since $\lim_{T \to 0} \theta_{\text{U}} = \pi$, there must exist a critical temperature $T_c$ at which the Uhlmann phase undergoes a sudden jump, indicating a temperature-induced topological phase transition.
For the simple case studied here, the critical temperature showing a jump of the Uhlmann phase emerges when $\cos(\pi \chi)=0$ or $\chi = 1/2$ since $0<\chi<1$ at finite temperatures, which yields $T_c\approx 0.75R$.

To visualize the transition, we plot $\theta_{\text{U}}$ as a function of $T$ on a semi-logarithmic scale in Figure~\ref{Fig1}. As shown in the plot, the Uhlmann phase indeed jumps by a quantized value of $\pi$. The quantized jump in the Uhlmann phase, which signifies a temperature-induced topological phase transition, indicates that below $T_c$, the final purification $W(\tau)$ is antiparallel to the initial purification $W(0)$ while above $T_c$, $W(\tau)$ becomes parallel to $W(0)$. Such a change of the topological property resembles that between a Mobius strip and a cylinder \cite{UP2D15}. Therefore, the temperature-induced topological phase transition shown in Figure~\ref{Fig1} has the Hamiltonian and eigenstates intact as temperature varies, so it is not a typical phase  or topological transition due to a change of thermodynamics or ground states but features a change of the topological object (the holonomy) in the underlying geometry of the mixed states.
We notice the number of critical temperatures where the Uhlmann phase exhibits quantized jumps may increase with the dimension of the Hilbert space, as illustrated in Ref.~\cite{OurPRA21}. However, the number of critical temperatures is not expected to increase in the presence of degeneracy.

\subsection{4D tight-binding model}
Next, we consider a more complex case of a four-band model that realizes a four-dimensional (4D) tight-binding Hamiltonian, originally introduced in Ref.~\cite{PhysRevB.78.195424}. The Hamiltonian in real space is a lattice model given by
\begin{align}\label{4Dtbmodel}
H=\sum_{n}\left[\sum_{j=1}^4\left(\psi^\dag_n\frac{\Gamma^0-\mi\Gamma^j}{2}\psi_{n+j}+\text{H.c.}\right)+m\psi^\dag_n\Gamma^0\psi_n\right].
\end{align}
Here $n$ and $j$ denote the lattice location and its four nearest-neighbors.
The real-space Hamiltonian can be transformed into the following expression in momentum space
\begin{align}
H=\sum_{\mathbf{k}}\psi^\dag_{\mathbf{k}}R_a\Gamma^a\psi_{\mathbf{k}}.
\end{align}
This corresponds to the four-level model of Eq.~\eqref{H} by identifying the components of the vector $\mathbf{R}$ as $R_1 = \sin k_x$, $R_2 = \sin k_y$, $R_3 = \sin k_z$, $R_4 = \sin k_u$, and $R_5 = m + \cos k_x + \cos k_y + \cos k_z + \cos k_u$. The parameter space is the first Brillouin zone with $k_x$, $k_y$, $k_z$, and $k_u$ spanning the interval $[-\pi, \pi]$, which forms a four-dimensional torus $T^4$.

For the 4D tight-binding model, the full expression of the Uhlmann connection is provided in Appendix \ref{app1}. We consider a special loop in $T^4$ defined by $C(t):=(k_x(t),k_y(t),k_z(t),k_u(t))=(k_x(t),0,0,0)$ and evolve the system accordingly. In this case, the Uhlmann connection reduces to
\begin{align}\label{AU4D}
\mathcal{A}_{\text{U}}=&- \frac{1-\operatorname{sech}(\beta R) }{2 R^2} \begin{pmatrix}
0 & 0 & 0 & -1 \\
0 & 0 & -1 & 0 \\
0 & 1 & 0 & 0 \\
1 & 0 & 0 & 0
\end{pmatrix}\notag\\&\times\left[ (m + 3) \cos k_x + 1\right] \dif k_x,
\end{align}
To evaluate the Uhlmann phase, we define \begin{align}\label{ImT}I(m,T):=\mathlarger{\int}_{0}^{2\pi} \frac{\operatorname{sech}(\beta R)-1}{2R^2}\left[ (m + 3) \cos k_x + 1\right] \dif k_x,
\end{align}
where $R=\sqrt{(m + 3)^2+ 2(m + 3) \cos k_x +1}$.
The Uhlmann holonomy in this case is given by
\begin{align}\label{holo4D}
\mathcal{P} \mathrm{e}^{-\oint_C \mathcal{A}_\text{U}}=\left(
	\begin{array}{cccc}
		\cos (I ) & 0 & 0 & \sin (I ) \\
		0 & \cos (I ) & \sin (I) & 0 \\
		0 & -\sin ( I ) & \cos (I ) & 0 \\
		-\sin ( I ) & 0 & 0 & \cos (I ) \\
	\end{array}
	\right).
\end{align}
Suppose the evolution starts at $k_x=0$, then the initial density matrix is
\begin{align}\label{rho04D}
\rho(0)=\frac{1}{4}(1_4-\tanh(\beta R)\Pi).
\end{align}
Here $\Pi=\operatorname{diag}(1,1,-1,-1)$.
 With the setup, we can numerically evaluate the Uhlmann phase given by
\begin{align}
\theta_{\text{U}}=\arg \operatorname{Tr}[\rho(0)\mathrm{e}^{-\oint_C \mathcal{A}_\text{U}}]=\arg (\cos (I(m,T))).
\end{align}

\begin{figure}[t]
	\centering
	\includegraphics[width=3.0in,clip]{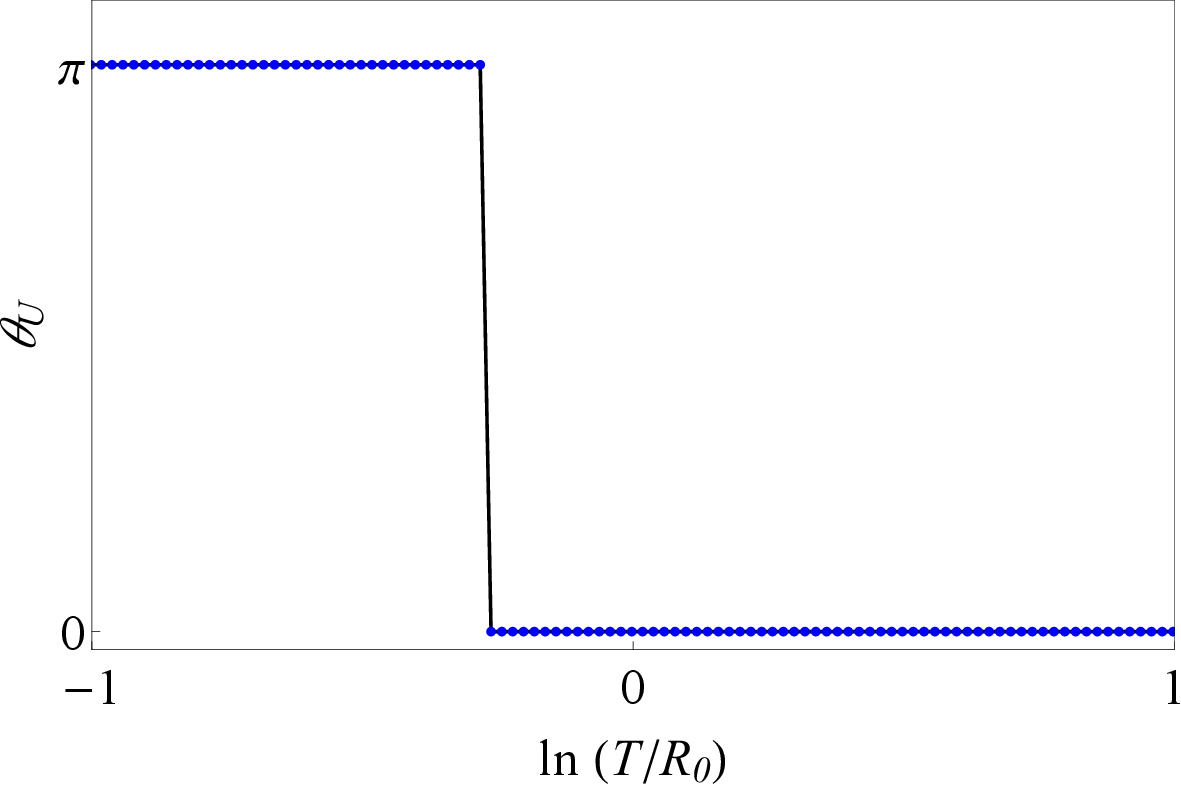}\\
	\includegraphics[width=3.5in,clip]{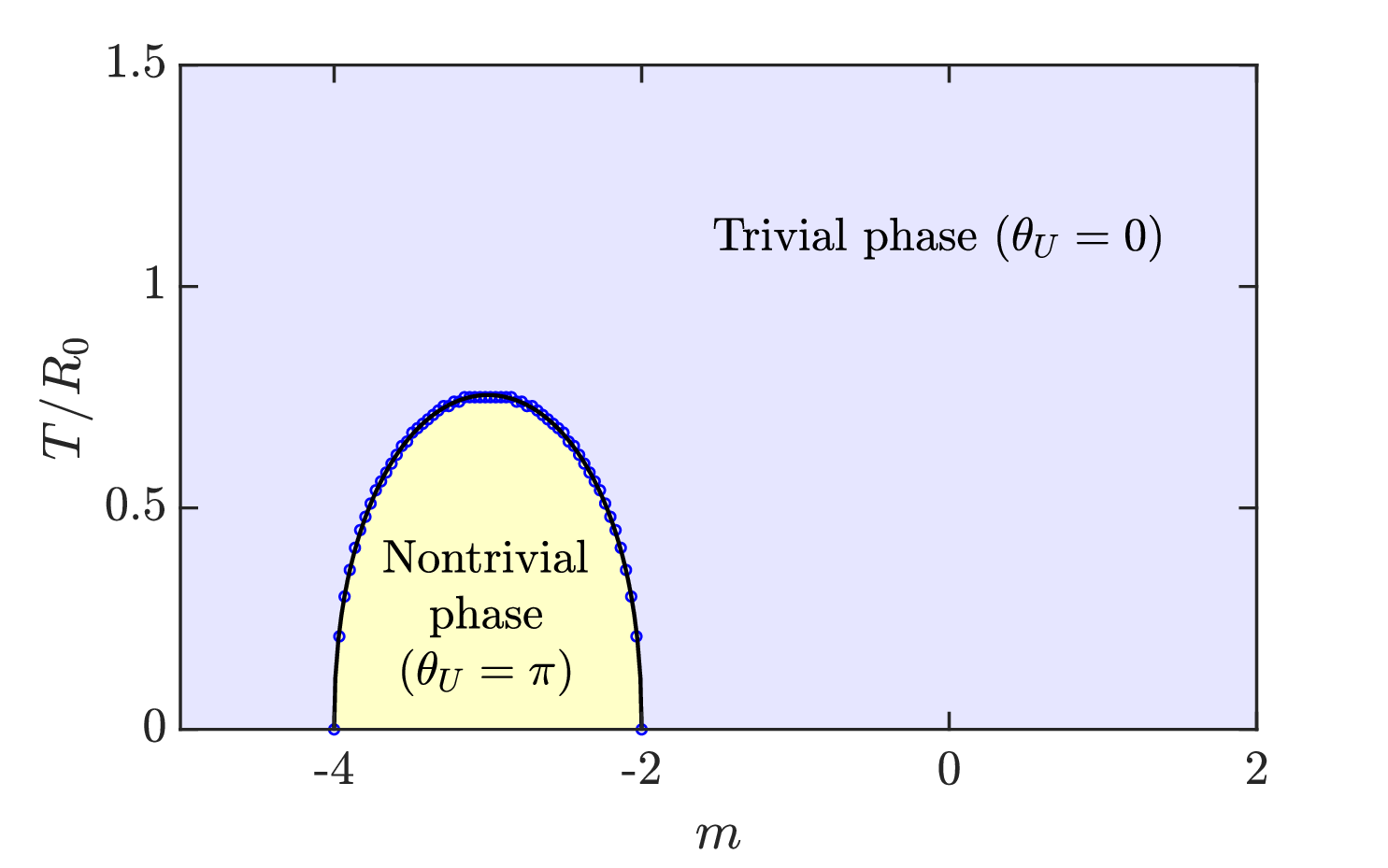}
	\caption{(Top panel) Uhlmann phase $ \theta_{\text{U}} $ as a function of temperature $ T $ on a semi-logarithmic scale for $ m = -3 $. A $ \pi $-jump occurs at $ T \approx 0.75R_0 $, signaling a temperature-induced topological phase transition.
(Bottom panel) Uhlmann-phase diagram of the 4D tight-binding model.
The numerical values of $ T_c $ are marked by the blue circles. The background and dome-shape regions correspond to the trivial ($ \theta_{\text{U}} = 0 $) and topological ($ \theta_{\text{U}} = \pi $) phases, respectively. The scalar WZ phase at $T=0$, however, vanishes after the system traverses the path $C$ regardless of $m$. Here $R_0=R(m=-3)$.
}
	\label{Fig2}
\end{figure}

In the top panel of Fig.~\ref{Fig2}, we plot the Uhlmann phase $ \theta_\text{U} $ as a function of temperature $ T $ on a semi-logarithmic scale for $ m = -3 $. A clear $ \pi $-jump is observed at the critical temperature $ T_c = 0.75R_0 $, where $ R_0 = R(m = -3) $. In the bottom panel, we present the Uhlmann-phase diagram of the 4D tight-binding model on the $ m\text{-}T $ plane. A finite Uhlmann phase appears exclusively in the range $ -4 < m < -2 $ at $T=0$ and forms a dome-shape region at finite temperatures. The rest of the diagram is occupied by the topologically trivial region with zero Uhlmann phase.
By fitting the boundary of the dome-shaped topological region with a symmetric power-law envelope function, we obtain an approximate relation between $ T_c $ and $ m $: $T_c /R_0\approx 0.75 \left[1 - (m + 3)^2\right]^{0.45}$.

For the calculation of the scalar WZ phase of the ground states of the 4D tight-binding model, we use Eqs.~(\ref{A+-ac12}) and (\ref{A+-ab12}) to find
\begin{align}\label{AWZ0}
\mathcal{A}_{-}=0
\end{align}
in this case. Consequently, the scalar WZ phase vanishes after the loop $C$ is traversed in the parameter space, regardless of the value of $m$. This explicit example demonstrates that the $T\rightarrow 0$ limit of $\theta_\text{U}$ does not always reduce to $\theta_\text{WZ}$, showcasing a fundamental difference from the previous case. As explained below Eq.~\eqref{Eq:sWZ}, the scalar WZ phase is a valid geometric phase of the degenerate ground states, not given simply by the $T\rightarrow 0$ limit of the Uhlmann phase. We will analyze and explain the disagreement in the 4D tight-binding model in later discussions.

\section{Correspondence between the Uhlmann and scalar WZ phases}\label{Sec4}
Here we first review the correspondence between the $T\rightarrow 0$ Uhlmann phase and Berry phase for non-degenerate systems. In an attempt to develop a similar correspondence, we found obstructions from degeneracy, symmetry, and topological properties exemplified by the 4D tight-binding model.

\subsection{Complications from degeneracy}
For generic non-degenerate cases, a conditional proof of the correspondence between the $T\rightarrow 0$ Uhlmann phase and the ground-state Berry phase has been given in Ref.~\cite{OurUB}. Appendix~\ref{app_corr} presents a more general proof with less stringent conditions.
The validity of the correspondence hinges on two conditions (adiabaticity and non-degeneracy) that are  violated in systems exhibiting Dirac points, which represent critical points in the parameter space where the energy gap of the system collapses (for example, $R\to 0$ in Eq.~\eqref{H}), resulting in degeneracy. These singularities emerge at energy-level crossings and may lead to non-analytic behavior in geometric observables.

These considerations render the correspondence between the Uhlmann phase in the zero-temperature limit and the Berry phase inapplicable to systems with degeneracy and Dirac points. We will present a more careful analysis and examine how topological obstructions, including Dirac points due to the Hamiltonian and zero-field axis due to symmetry in the 4D tight-binding model, prevent an agreement between the Uhlmann phase in the zero-temperature limit and the geometric phase of the ground state when degeneracy is present.

\subsection{Correspondence between Uhlmann and WZ phases is not guaranteed}\label{Sec4.1}
The two examples discussed in Sec.~\ref{Sec3} sharply contrast different outcomes regarding the relationships between the $T\rightarrow 0$ Uhlmann phase and the scalar WZ phase. We now turn to a more general analysis of this issue from an analytic perspective. Our objectives are twofold: to determine whether a simple relationship exists between the Uhlmann connection in the zero-temperature limit and the WZ connection and
to identify the conditions under which the Uhlmann phase in the zero-temperature limit may reduce to the scalar WZ phase. In the following, we consider generic systems with $N$-dimensional Hilbert space.

In a system where the $i$-th eigenstate $|\psi^{(i)}_n\rangle$ has degeneracy $D_i$, with $n = 1, \dots, D_i$ and $\sum_i D_i = N$, the Uhlmann connection $\mathcal{A}_\text{U}$ in Eq.~(\ref{AUE2}) can be more precisely formulated as:
\begin{align}\label{AUT0}
\mathcal{A}_\text{U} =& -\sum_{i \neq j} \sum_{n=1}^{D_i} \sum_{m=1}^{D_j} \frac{\left( \sqrt{\lambda_i} - \sqrt{\lambda_j} \right)^2}{\lambda_i + \lambda_j} \notag\\&\times |\psi^{(i)}_n\rangle \langle \psi^{(i)}_n| \dif \psi^{(j)}_m\rangle \langle \psi^{(j)}_m|,
\end{align}
where $|\dif \psi^{(j)}_m\rangle \equiv \dif |\psi^{(j)}_m\rangle$. Here $\lambda_n = \frac{e^{-\beta E_n}}{Z}$ with the energy levels ordered as $E_0 < E_1 < \dots$. Since $\lambda_i \neq \lambda_j$, we define $\lambda_m = \min\{\lambda_i, \lambda_j\}$ and $\lambda_M = \max\{\lambda_i, \lambda_j\}$. In the low-temperature limit, we find,
\begin{align}
\lim_{T \to 0} \frac{\left( \sqrt{\lambda_i} - \sqrt{\lambda_j} \right)^2}{\lambda_i + \lambda_j} = \lim_{T \to 0} \frac{\left( 1 - \sqrt{\frac{\lambda_m}{\lambda_M}} \right)^2}{1 + \frac{\lambda_m}{\lambda_M}} = 1.
\end{align}
When $T\to 0$, the Uhlmann connection can be expressed as
\begin{align}\label{AU}
\mathcal{A}_\text{U} \to& -\sum_{i \neq j} \sum_{n=1}^{D_i} \sum_{m=1}^{D_j} |\psi^{(i)}_n\rangle \langle \psi^{(i)}_n| \dif \psi^{(j)}_m\rangle \langle \psi^{(j)}_m|\notag\\
=&-\sum_{i, j} \sum_{n=1}^{D_i} \sum_{m=1}^{D_j} |\psi^{(i)}_n\rangle \langle \psi^{(i)}_n| \dif \psi^{(j)}_m\rangle \langle \psi^{(j)}_m|\notag\\
&+\sum_{i} \sum_{n,m=1}^{D_i} |\psi^{(i)}_n\rangle \langle \psi^{(i)}_n| \dif \psi^{(i)}_m\rangle \langle \psi^{(i)}_m|\notag\\
=&-\sum_{j,m}| \dif \psi^{(j)}_m\rangle \langle \psi^{(j)}_m|+\mathcal{A}_\text{WZ},
\end{align}
where
\begin{align}
\mathcal{A}_\text{WZ}=&\sum_{i} \sum_{n,m=1}^{D_i}\langle \psi^{(i)}_n| \dif \psi^{(i)}_m\rangle |\psi^{(i)}_n\rangle  \langle \psi^{(i)}_m|\notag\\
=&\begin{pmatrix}\mathcal{A}^{(0)}_\text{WZ} & &\\ & \mathcal{A}^{(1)}_\text{WZ} & \\ & & \ddots
\end{pmatrix}
\end{align}
is the full WZ connection matrix. Here we have suppressed the explicit dependence on the parameter $\mathbf{R}$ for brevity. According to Eq.~(\ref{AU}), the Uhlmann holonomy along a closed loop $C$ in the parameter space is expressed as:
\begin{align}\label{holoToy}
\lim_{T \to 0} \mathcal{P}\me^{- \mathlarger{\oint}_C \mathcal{A}_\text{U} } = \mathcal{P}\me^{ - \mathlarger{\oint}_C \left( \mathcal{A}_\text{WZ} - \sum_{j,m} | \dif \psi^{(j)}_m \rangle \langle \psi^{(j)}_m | \right) }.
\end{align}

In the first example above, when the system evolves along the equator, the eigenvalues $\pm R$ remain constant, and the associated evolution corresponds to a unitary transformation of the states involved. This can be written as $|\psi^{(j)}_m(\mathbf{R}(t))\rangle = \mathcal{D}(\mathbf{R}(t)) |\psi^{(j)}_m(\mathbf{R}(0))\rangle$. For notational simplicity, we denote $|\psi^{(j)}_m(0)\rangle = |\psi^{(j)}_m(\mathbf{R}(0))\rangle$. In this scenario, the relationship between $\mathcal{A}_\text{U}$ in the $T\rightarrow 0$ limit and $\mathcal{A}_\text{WZ}$ simplifies to
\begin{align}\label{AU2}
\mathcal{A}_\text{U} \to &  \mathcal{A}_\text{WZ}-\sum_{j,m} \dif \mathcal{D} | \psi^{(j)}_m(0) \rangle \langle \psi^{(j)}_m(0) | \mathcal{D}^\dag  \notag \\
& =  \mathcal{A}_\text{WZ}-\dif \mathcal{D} \mathcal{D}^\dag.
\end{align}
Thus, we obtain a direct relation that explicit connects the Uhlmann connection and the Wilczek-Zee connection.
Moreover, the relation (\ref{holoToy}) simplifies to
\begin{align}\label{holoToy2}
\lim_{T \to 0} \mathcal{P}\me^{- \mathlarger{\oint}_C \mathcal{A}_\text{U} } = \mathcal{P}\me^{ - \mathlarger{\oint}_C \left( \mathcal{A}_\text{WZ} -  \dif \mathcal{D} \mathcal{D}^\dag\right) }.
\end{align}
Assuming the additional condition $[\mathcal{A}_\text{WZ} , \dif \mathcal{D} \mathcal{D}^\dag]=0$ holds, the expression further reduces to
\begin{align}\label{holoToy3}
\lim_{T \to 0} \mathcal{P}\me^{- \mathlarger{\oint}_C \mathcal{A}_\text{U} } = \left(\mathcal{P}\me^{ - \mathlarger{\oint}_C \mathcal{A}_\text{WZ}  } \right)\left(\mathcal{P}\me^{ \mathlarger{\oint}_C   \dif \mathcal{D} \mathcal{D}^\dag}\right).
\end{align}

Since $\dif \mathcal{D} \mathcal{D}^\dag$ is anti-Hermitian, the path-ordered exponentiation $\mathcal{P}\me^{ \mathlarger{\oint}_C   \dif \mathcal{D} \mathcal{D}^\dag}$ forms an $N \times N$ unitary matrix. Given that this path-ordered exponentiation maps the closed loop $C \sim S^1$ to the group $U(N)$ of unitary matrices, the resulting $\mathcal{P}\me^{ \mathlarger{\oint}_C   \dif \mathcal{D} \mathcal{D}^\dag}$ is an element of the fundamental group $\pi_1(U(N)) = \mathbb{Z}$. Moreover, since $\dif \mathcal{D} \mathcal{D}^\dag$ is the Maurer-Cartan form of the $U(N)$ group, we then obtain~\cite{AGbook}
\begin{align}
\mathcal{P} \exp \left( \mathlarger{\oint}_C \dif \mathcal{D} \mathcal{D}^\dag \right) = \me^{2\pi \mi \kappa} 1_N,
\end{align}
where $1_N$ denotes the $N \times N$ identity matrix, and the winding number $\kappa$ associated with $\mathcal{D}$ along the loop $C$ is given by:
\begin{align}
\kappa = \frac{1}{2\pi \mi} \oint_C \text{Tr} \left( \dif \mathcal{D} \mathcal{D}^\dag \right).
\end{align}
If $C$ is a loop contractible to a point, then $\kappa = 0$. In this case,
\begin{align}
\lim_{T \to 0} \mathcal{P}\me^{- \mathlarger{\oint}_C \mathcal{A}_\text{U} } = \me^{2\pi \mi k}\mathcal{P}\me^{ - \mathlarger{\oint}_C \mathcal{A}_\text{WZ}  } .
\end{align}
Moreover, in the low temperature limit,
\begin{align}
\rho\to \frac{1}{D_0}\sum_{n}|\psi^{(0)}_n\rangle\langle\psi^{(0)}_n|=
\begin{pmatrix}\frac{1}{D_0}P_0 &  & \\ & 0  & \\ & &\ddots,
\end{pmatrix},
\end{align}
where $P_0$ is the projector to the ground state with degeneracy $D_0$. The $T\rightarrow 0$ Uhlmann phase thus becomes
\begin{align}
\theta_\text{U}(C)&\to\arg\text{Tr}_-\left(\frac{1}{D_0}P_0 \me^{2\pi \mi \kappa}\mathcal{P}\me^{ - \mathlarger{\oint}_C \mathcal{A}^{(0)}_\text{WZ}  } \right)\notag\\
&=\theta_\text{WZ}(C)+2\pi \kappa \notag\\
&=\theta_\text{WZ}(C) \mod 2\pi.
\end{align}

The above argument establishes a conditional relation between the Uhlmann phase as $T\to 0$ and the scalar WZ phase when the system evolves unitarily under a transformation $\mathcal{D}$, provided the condition $[\mathcal{A}_\text{WZ} , \dif \mathcal{D} \mathcal{D}^\dag]=0$ holds. However, this condition is somewhat too restricted. In fact, the first example we presented above (the simple case) does not satisfy this commutation relation, yet the correspondence between the $T\rightarrow 0$ Uhlmann phase and the scalar WZ phase remains valid. This suggests that the actual requirement for the correspondence to hold may be more lenient, though a rigorous mathematical characterization still awaits future research.
In contrast, the second example of the 4D tight-binding model shows non-unitary evolution since the eigenvalues vary as the system evolves in the parameter space. Therefore, the relationship between the $T\rightarrow 0$ Uhlmann phase and the scalar WZ phase becomes considerably more subtle and dependent on specific features of the system. Our results thus illustrate the complex physics facing finite-temperature topological quantum systems.

\section{Obstructions to Phase Relations}\label{Sec:PhaseReduction}
\subsection{ Zero-field axis: Geometric Origins and Physical Implications}
To illustrate the impact of Dirac points on the zero-temperature limit of the Uhlmann phase, we contrast the two examples analyzed so far. In the first case, the system evolves unitarily since the eigenvalues of the Hamiltonian remain constant along the evolution path, thereby yielding an agreement between the $ \lim_{T\to 0} \theta_{\text{U}} $ and $ \theta_{\text{WZ}} $. By contrast, in the 4D tight-binding model along the curve $ C(t) = (k_x(t), 0, 0, 0) $, the behavior near the Dirac points at $ (m, k_x) = (-4, 0) $ and $ (-2, \pi) $, where $R=\sqrt{(m + 3)^2+ 2(m + 3) \cos k_x +1}=0$, reveals parameter-dependent topology. The WZ connection vanishes under the constraints $ R_2 = R_3 = R_4 = 0 $ and $ \dif R_2 = \dif R_3 = \dif R_4 = 0 $ as shown in Eq.~\eqref{AWZ0}, resulting in $ \theta_{\text{WZ}} = 0 $. This behavior actually stems from the specific geometry of $U(2)$ monopoles located at Dirac points: The WZ connection is a non-abelian $U(2)$ gauge potential, qualitatively different from the  magnetic field generated by $U(1)$ monopoles.

Since the WZ connection (playing the role of a gauge potential) vanishes along $C(t)$, we denote the path as a zero-field axis. We remark that $k_x$ itself is a line, but the periodic boundary condition makes it topologically equivalent to a loop. This vanishing of the WZ connection is because the zero-field axis resides in a high-symmetry subspace $\mathcal{S}$ invariant under $SO(2) \simeq U(1)$ transformations corresponding to the Cartan subalgebra of $U(2)$, where the vanishing Hamiltonian components ($R_2 = R_3 = R_4 = 0$) freeze the geometric evolution. We mention that similar zero-field behavior has been observed in quantum systems featuring twistorial monopoles \cite{JHEP082023088}.

The appearance of a zero-field axis is governed by two conditions detailed in in Appendix \ref{app5}. First, the parameter subspace $\mathcal{S}$ must be invariant under a continuous symmetry subgroup $\mathcal{H}\subset G$, keeping the path rigid under any transformation. Second, $\mathcal{S}$ must align with the Cartan subalgebra of $G$, whose maximal Abelian structure permits a cancellation of all non-Abelian gauge components. Together these constraints impose wave-function rigidity along the path, resulting in $\langle \psi^{(0)}_n | \partial_\mu \psi^{(0)}_m \rangle = 0$ along the degenerate subspace as the system evolves along $C(t)$. For a differential geometry proof, see Appendix \ref{app5}. The symmetry subgroup $\mathcal{H}$ freezes the internal state evolution, while the Cartan alignment reduces the connection to mutually commuting fields that annihilate.

In the 4D tight-binding model, the $SO(2)\simeq U(1)$ symmetry satisfies both symmetry invariance and Cartan alignment  (see Appendix \ref{app5}), thereby establishing the path $C(t)$ as a zero-field axis where the connection vanishes. The zero connection occurs even though the path passes through monopole singularities from the Dirac points. Interestingly, the effect is topology-independent because the symmetry protection guarantees zero field-strength along the entire axis whether it is open or closed. Therefore, the example offers a possible realization of a zero-field axis for non-Abelian gauge potentials arising from $U(N)$ ($N\ge 2$) effective monopoles in condensed-matter systems.

By contrast, the Uhlmann connection and the Uhlmann phase in the same example are not affected by the symmetry of the path in such a fashion. From Eq. (\ref{AUT0}), it is evident that the Uhlmann connection depends only on the coupling between different energy levels since the right hand side of Eq. (\ref{AUT0}) has finite contributions only when $i \neq j$. Consequently, the Uhlmann connection becomes less sensitive to degeneracy and, in turn, to a symmetry axis within a degenerate subspace. Moreover, the zero-temperature limit of the Uhlmann phase is strongly influenced by the Dirac points, unlike the scalar WZ phase in this case. The contrast thus highlights the geometric distinction between mixed and pure states in topological systems.

\subsection{Singular behavior of the Uhlmann connection and phase at Dirac points}
The following asymptotic analysis of the 4D tight-binding model elucidates the singular behavior of the Uhlmann phase at the Dirac points. From Eq.~(\ref{AU4D}), the kernel function
\begin{align}\label{kernelf}
f(k_x, T) = -\frac{1 - \operatorname{sech}(\beta R)}{2 R^2} \left[ (m + 3) \cos k_x + 1 \right]
\end{align}
governs the Uhlmann connection. Near the $ m = -4 $ Dirac point with $ k_x = q \to 0 $, $R \approx |q|$ and
\begin{align}\label{f0}
f(q, T) \approx \frac{\operatorname{sech}(\beta |q|) - 1}{4}.
\end{align}
Similarly, near the $ m = -2 $ Dirac point with $ k_x = \pi + q \to \pi $, $R \approx |q|$ and
\begin{align}
f(\pi+q, T) \approx \frac{\operatorname{sech}(\beta |q|) - 1}{4}.
\end{align}
The critical scaling $\frac{\operatorname{sech}(\beta |q|) - 1}{4} \sim -\frac{1}{4 T^2} $ for $|q| \sim T$
shows divergent behavior as $T\rightarrow 0$.


At the $ m = -4 $ Dirac point, the Uhlmann phase exhibits critical dependence on the order of limits. In contrast, the scalar WZ phase remains zero around the critical point due to the zero-field axis. When approaching from the topological phase ($ m > -4 $) by first taking the zero-temperature limit, the Uhlmann phase maintains its quantized value:
\begin{align}
\lim_{m \to -4^+} \left( \lim_{T \to 0} \theta_{\text{U}}(m) \right) = \pi.
\end{align}
Some details can be found in Appendix \ref{app4}.
However, a direct evaluation at $ m = -4 $ reveals singular behavior as the integral
\begin{align}
I(-4,T) &= \int_0^{2\pi} \frac{\operatorname{sech}(\beta R)-1}{2R^2} \left[ - \cos k_x + 1 \right]  \dif k_x,\notag \\
R &= 2 \left| \sin \frac{k_x}{2} \right| \approx |k_x| \quad \text{near} \quad k_x=0
\end{align}
becomes divergent near $ k_x = 0 $ since the integrand scales as $ f(k_x,T) \sim -1/(4T^2) $ when $ |k_x| \sim T $. Therefore,
\begin{align}
\lim_{T \to 0} I(-4,T) = -\infty,
\end{align}
rendering the zero-temperature Uhlmann phase undefined on the critical point:
\begin{align}
\lim_{T \to 0} \theta_{\text{U}}(-4) = \lim_{T \to 0} \arg \left( \cos I(-4,T) \right) \quad \text{undefined,}
\end{align}
whereas approaching from the trivial phase ($ m < -4 $) yields (see Appendix \ref{app4}):
\begin{align}
\lim_{m \to -4^-} \left( \lim_{T \to 0} \theta_{\text{U}}(m) \right) = 0.
\end{align}
Therefore, the $T\rightarrow 0$ Uhlmann phase exhibits a quantized jump across a Dirac point, exemplified by the 4D tight-binding model.
The explicit dependence on taking the limit of the Uhlmann phase originates from the Dirac point, where the gap closure induces divergent behavior in the Uhlmann connection. This is in stark contrast with the scalar WZ phase exhibiting continuous behavior. Therefore, the degeneracy, zero-field axis, and Dirac points can hinder a universal correspondence between the Uhlmann phase in the zero-temperature limit and the scalar WZ phase at zero temperature.

\section{Implications}\label{Sec5}
Degenerate systems have been of interest in holonomic quantum computation (HQC)~\cite{Pachos12,ZHANG20231,10.21468/SciPostPhysCore.3.2.014}, where the phase factors are engineered to achieve quantum gate operations. Some exemplary systems considered in this direction include Majorana fermions and atoms interacting with laser light. The WZ connection contributes to the non-Abelian holonomy, which is frequently used in various schemes of HQC. The introduction of the scalar WZ phase will help further quantify the holonomy induced by the WZ connection int the degenerate ground-state subspace. Moreover, quantum geometries of degenerate pure states have been characterized~\cite{Avdoshkin24,PhysRevB.81.245129,PhysRevA.109.043305}, which may inspire future research to bridge local geometry and global topology of degenerate quantum systems.

Since a typical Uhlmann process is incompatible with the Hamiltonian dynamics~\cite{Guo20}, measuring the Uhlmann phase in natural systems typically encounter the uncontrollable problem of the environment to satisfy the Uhlmann parallel-transport condition. Nevertheless, the Uhlmann phase of two-level \cite{npj18} and three-level~\cite{Mastandrea25} systems have been simulated on the IBM quantum platform because the ancilla in the simulation modeling the environment can be fully controlled to respect the Uhlmann parallel-transport condition. Further, there is a proposal for generalizing the scheme to spin-$j$ systems~\cite{OurPRA21}. The four-level model analyzed in this work may be encoded by two qubits, which will require another two qubits to serve as the ancilla to enact the system-environment coupling for generating the Uhlmann phase. Although simulating the Uhlmann phase of degenerate systems on quantum computers can be a challenging task, its success will allow us to explore the rich physics behind finite-temperature topological systems.

The comparison between the $T\rightarrow 0$ results from the Uhlmann connection and those from the Berry or WZ connection of the corresponding pure states helps address the fundamental issue that although the density matrix smoothly transit from the $T\rightarrow 0$ mixed state to the $T=0$ pure state, the Uhlmann fiber bundle is for full-rank density matrices but the pure-state bundles only concerns the ground-state subspace. Therefore, any correspondence between the $T\rightarrow 0$ results and the pure-state results needs to be analyzed on a case-by-case basis~\cite{OurUB}. The conditional correspondence between the $T\rightarrow 0$ Uhlmann phase and the scalar WZ phase derived here may serve as a starting point for developing a broader classification of the correspondence between the two phases in degenerate quantum systems.

Interesting time-reversal invariant (TRI) topological insulators can exist in 4D, which are described by the effective Chern-Simons theory \cite{zhang2001four,bernevig2002effective}. The 4D tight-binding model in Eq.~\eqref{4Dtbmodel} has the identical form of the generic 4D model of Ref.~\cite{PhysRevB.78.195424} in the discussion of 4D TRI insulators, which may be realized experimentally by 2D quasi-periodic lattices with twisted boundary conditions in both directions~\cite{PhysRevLett.111.226401}.
Moreover, through the method of dimension reduction~\cite{PhysRevB.78.195424}, both 2D TRI quantum spin Hall (QSH) insulator \cite{kane2005quantum,bernevig2006quantum} and 3D topological insulator \cite{bernevig2006quantum1,konig2007quantum,fu2007topological,moore2007topological,roy2009z,murakami2004spin} can be derived from their parent systems in 4D, yielding possibly relations and classifications.
The effective topological field theory of TRI insulators predicts measurable phenomena, most strikingly the topological magnetoelectric effect where an electric field generates a magnetization in the same direction with a universal quantized coefficient~\cite{niemi1983axial}
Therefore, the properties and distinctions of the scalar WZ phase and the Uhlmann phase discussed above serve as a potential theoretical framework for future investigations of 4D TRI topological insulators with degeneracy or Dirac points at zero or finite temperature.

We mention that alternative approaches to mixed-state geometric phases have been proposed, including the interferometric geometric phase (IGP) \cite{PhysRevLett.85.2845, PhysRevA.67.020101, PhysRevA.70.052109, Faria_2003, Chaturvedi2004, PhysRevLett.93.080405, Kwek_2006} inspired by optical interferometry. The IGP provides an experimentally accessible phase observable without requiring cyclic evolution. Moreover, it can be shown that the IGP is equivalent to a weighted sum of the Berry phases over the eigenstates of the system for cyclic processes~\cite{PhysRevLett.85.2845,PhysRevA.68.022106}.
The experimental realizations of the IGP in platforms such as NMR and polarized neutron systems \cite{PhysRevLett.91.100403, Ghosh_2006, PhysRevLett.101.150404, PhysRevLett.94.050401} underscore its practical value. There have been some attempts to generalize the IGP to degenerate systems~\cite{PhysRevA.67.032106}. However, the underlying parallel-transport condition of the IGP still awaits a concise fiber-bundle foundation to characterize its fully geometric meaning \cite{PhysRevB.107.165415}.

\section{conclusion}\label{Sec6}
We have investigated the mixed-state Uhlmann phase of an exemplary degenerate system and searched for possible relationships with the WZ holonomy in the degenerate ground-state subspace through the introduction of the scalar WZ phase. The four-level model featuring two doubly degenerate subspaces allows for explicit forms of the Uhlmann and WZ connections to evaluate their holonomy and phases. Through explicit examples, we show that the Uhlmann phase in the zero-temperature limit may or may not approach the scalar WZ phase of the degenerate ground states. Moreover, our theoretical analysis gives some conditions under which the agreement between the zero-temperature limit of the Uhlmann phase and the scalar WZ phase is guaranteed. Furthermore, profound topological obstructions, such as the zero-field axis and Dirac points discussed in the 4D tight-binding model, may account for the disagreement between the two phases.

The rich phenomena behind the mixed-state Uhlmann phase and pure-state scalar WZ phases of systems with degeneracy highlight challenges facing geometric and topological properties of degenerate quantum systems at and beyond zero temperature. Our examples and analyses make some progress towards understanding the deeper connection between mixed-state and pure-state topology in the presence of degeneracy and offer inspiration for investigating temperature effects in topological systems with potential applications in condensed matter physics and quantum information science.

\begin{acknowledgments}
H.G. was supported by the Innovation Program for Quantum Science and Technology (Grant No. 2021ZD0301904) and the National Natural Science Foundation of China (Grant No. 12074064). C.C.C. was supported by the NSF (No. PHY-2310656) and DOE (No. DE-SC0025809).
\end{acknowledgments}

\appendix
\section{More details of the Uhlmann connection}\label{app1}
Using the diagonal form $\sqrt{\rho} = \sum_i \sqrt{\lambda_i} |i\rangle \langle i|$, we obtain
\begin{align}\label{AUTl3}
[\sqrt{\rho}, \dif \sqrt{\rho}] =& \sum_{i,j} \sqrt{\lambda_i \lambda_j} \left( |i\rangle \langle i| \dif |j\rangle \langle j| - |j\rangle (\dif \langle j|) |i\rangle \langle i| \right) \notag\\
+& \sum_i \lambda_i \left( |i\rangle \dif \langle i| - \dif |i\rangle \langle i| \right).
\end{align}
By exchanging indices $i \leftrightarrow j$ in the second term of the first line on the right-hand side and using $(\dif \langle i|) |j\rangle = - \langle i| \dif |j\rangle$, the first line becomes
$2 \sum_{i,j} \sqrt{\lambda_i \lambda_j} |i\rangle \langle i| \dif |j\rangle \langle j|$.
Following a similar derivation, the second line becomes
$\sum_{i,j} \lambda_i \left( |i\rangle (\dif \langle i|) |j\rangle \langle j| - |j\rangle \langle j| \dif |i\rangle \langle i| \right) = - \sum_{i,j} (\lambda_i + \lambda_j) |i\rangle \langle i| \dif |j\rangle \langle j|$.
Combining the above results, we obtain
\begin{align}\label{AUTl4}
[\sqrt{\rho}, \dif \sqrt{\rho}] = - \sum_{i,j} \left( \sqrt{\lambda_i} - \sqrt{\lambda_j} \right)^2 |i\rangle \langle i| \dif |j\rangle \langle j|.
\end{align}
Substituting this into Eq. (\ref{AUE}), we obtain an alternative expression for the Uhlmann connection:
\begin{align}\label{AUTlf}
\mathcal{A}_\text{U} =& - \sum_{i \neq j} \frac{\left( \sqrt{\lambda_i} - \sqrt{\lambda_j} \right)^2}{\lambda_i + \lambda_j} |i\rangle \langle i| \dif |j\rangle \langle j|\notag\\
 =& - \sum_{i \neq j} \left(1-\frac{2\sqrt{\lambda_i\lambda_j}}{\lambda_i+\lambda_j}\right) |i\rangle \langle i| \dif \left(|j\rangle \langle j|\right),
\end{align}
where the second line comes from the fact that $\langle i|j\rangle=0$ if $\lambda_i\neq\lambda_j$ in thermal equilibrium states. For the four-level model with two doubly degenerate spaces considered in the main text, $\lambda_\pm=\frac{\me^{\mp\beta R}}{4\cosh(\beta R)}$, satisfying $\lambda_++\lambda_-=\frac{1}{2}$ and $\sqrt{\lambda_+\lambda_-}=\frac{1}{4\cosh(\beta R)}$. Substituting the above expressions into Eq.~(\ref{AUTlf}) and plugging in $|i\rangle,|j\rangle=\{|\psi_a\rangle,|\psi_b\rangle,|\psi_c\rangle,|\psi_d\rangle\}$, we get
\begin{align}\label{M1b}
\mathcal{A}_{\text{U}} =-\left[1-\text{sech}\left(\frac{R}{T}\right)\right]\left(P_+\dif P_-+P_-\dif P_+\right),
\end{align}
which then leads to Eq.~\eqref{M1}.

For the 4D tight-binding model, the Uhlmann connection according to Eq.~(\ref{M1}) is give by
\begin{align}
\mathcal{A}_{\text{U}} =- \frac{1-\operatorname{sech}(\beta R) }{2 R^2} M,
\end{align}
where the elements of the matrix $ M $ are:
\begin{widetext}
\begin{align}
\begin{array}{rcl}
M_{11} &=& \mi \left[ \sin k_y \cos k_x \dif k_x - \sin k_x \cos k_y \dif k_y + \sin k_u \cos k_z \dif k_z - \sin k_z \cos k_u \dif k_u \right], \\
M_{12} &=& (\sin k_x - \mi \sin k_y) \left[ \cos k_z \dif k_z - \mi \cos k_u \dif k_u \right] - (\sin k_z - \mi \sin k_u) \left[ \cos k_x \dif k_x - \mi \cos k_y \dif k_y \right], \\
M_{13} &=& -\left( m + \sum_{\alpha} \cos k_\alpha \right) \cos k_z \dif k_z + \mi \left( m + \sum_{\alpha} \cos k_\alpha \right) \cos k_u \dif k_u \\
&&\quad + (\sin k_z - \mi \sin k_u) \left[ -\sin k_x \dif k_x - \sin k_y \dif k_y - \sin k_z \dif k_z - \sin k_u \dif k_u \right], \\
M_{14} &=& -\left( m + \sum_{\alpha} \cos k_\alpha \right) \cos k_x \dif k_x + \mi \left( m + \sum_{\alpha} \cos k_\alpha \right) \cos k_y \dif k_y \\
&&\quad + (\sin k_x - \mi \sin k_y) \left[ -\sin k_x \dif k_x - \sin k_y \dif k_y - \sin k_z \dif k_z - \sin k_u \dif k_u \right], \\
M_{21} &=& (\sin k_z + \mi \sin k_u) \left[ \cos k_x \dif k_x + \mi \cos k_y \dif k_y \right] - (\sin k_x + \mi \sin k_y) \left[ \cos k_z \dif k_z + \mi \cos k_u \dif k_u \right], \\
M_{22} &=& \mi \left[ -\sin k_y \cos k_x \dif k_x + \sin k_x \cos k_y \dif k_y - \sin k_u \cos k_z \dif k_z + \sin k_z \cos k_u \dif k_u \right], \\
M_{23} &=& (\sin k_x + \mi \sin k_y) \left[ -\sin k_x \dif k_x - \sin k_y \dif k_y - \sin k_z \dif k_z - \sin k_u \dif k_u \right] \\
&&\quad - \left( m + \sum_{\alpha} \cos k_\alpha \right) \left[ \cos k_x \dif k_x + \mi \cos k_y \dif k_y \right], \\
M_{24} &=& \left( m + \sum_{\alpha} \cos k_\alpha \right) \cos k_z \dif k_z + \mi \left( m + \sum_{\alpha} \cos k_\alpha \right) \cos k_u \dif k_u \\
&&\quad - (\sin k_z + \mi \sin k_u) \left[ -\sin k_x \dif k_x - \sin k_y \dif k_y - \sin k_z \dif k_z - \sin k_u \dif k_u \right], \\
M_{31} &=& M_{24}, \\
M_{32} &=& \left( m + \sum_{\alpha} \cos k_\alpha \right) \cos k_x \dif k_x - \mi \left( m + \sum_{\alpha} \cos k_\alpha \right) \cos k_y \dif k_y \\
&&\quad - (\sin k_x - \mi \sin k_y) \left[ -\sin k_x \dif k_x - \sin k_y \dif k_y - \sin k_z \dif k_z - \sin k_u \dif k_u \right], \\
M_{33} &=& \mi \left[ \sin k_y \cos k_x \dif k_x - \sin k_x \cos k_y \dif k_y - \sin k_u \cos k_z \dif k_z + \sin k_z \cos k_u \dif k_u \right], \\
M_{34} &=& (\sin k_x - \mi \sin k_y) \left[ \cos k_z \dif k_z + \mi \cos k_u \dif k_u \right] - (\sin k_z + \mi \sin k_u) \left[ \cos k_x \dif k_x - \mi \cos k_y \dif k_y \right], \\
M_{41} &=& \left( m + \sum_{\alpha} \cos k_\alpha \right) \cos k_x \dif k_x + \mi \left( m + \sum_{\alpha} \cos k_\alpha \right) \cos k_y \dif k_y \\
&&\quad - (\sin k_x + \mi \sin k_y) \left[ -\sin k_x \dif k_x - \sin k_y \dif k_y - \sin k_z \dif k_z - \sin k_u \dif k_u \right], \\
M_{42} &=& -\left( m + \sum_{\alpha} \cos k_\alpha \right) \cos k_z \dif k_z + \mi \left( m + \sum_{\alpha} \cos k_\alpha \right) \cos k_u \dif k_u \\
&&\quad + (\sin k_z - \mi \sin k_u) \left[ -\sin k_x \dif k_x - \sin k_y \dif k_y - \sin k_z \dif k_z - \sin k_u \dif k_u \right], \\
M_{43} &=& (\sin k_z - \mi \sin k_u) \left[ \cos k_x \dif k_x + \mi \cos k_y \dif k_y \right] - (\sin k_x + \mi \sin k_y) \left[ \cos k_z \dif k_z - \mi \cos k_u \dif k_u \right], \\
M_{44} &=& \mi \left[ -\sin k_y \cos k_x \dif k_x + \sin k_x \cos k_y \dif k_y + \sin k_u \cos k_z \dif k_z - \sin k_z \cos k_u \dif k_u \right].
\end{array}
\end{align}
\end{widetext}

\section{More details about the scalar WZ phase}\label{app2a}
We demonstrate the equivalence between the two expressions for the scalar WZ phase given in Eqs.~(\ref{Eq:sWZ}) and (\ref{thetaWZ}).
Using the purification $W_- = \frac{1}{\sqrt{D}} \sqrt{P_-} \mathcal{U}$, the initial and final states are expressed as
\begin{align}
W_-(0)=&\frac{1}{\sqrt{D}}\sum_{a=1}^D|\psi_a(0)\rangle\langle\psi_a(0)|\mathcal{U}(0),\notag\\
W_-(\tau)=&\frac{1}{\sqrt{D}}\sum_{a=1}^D|\psi_a(\tau)\rangle\langle\psi_a(\tau)|\mathcal{U}(\tau),
\end{align}
where $\mathcal{U}(\tau)$ and $\mathcal{U}(0)$ are related by Eq.~(\ref{Eq:WZholonomy}). Using these results and Eq.~(\ref{WZc}), the scalar WZ phase in Eq.~(\ref{Eq:sWZ}) can then be evaluated as:
\begin{align}\label{Eq:sWZ2}
&\theta_{\text{WZ}}(C) =\arg  \langle W_-(0) | W_-(\tau) \rangle=\arg\text{Tr}_-(W^\dag(0)W(\tau))\notag\\
=&\arg\sum_{a,b,c,d}\text{Tr}\big[\mathcal{U}^\dag(0)|\psi_a(0)\rangle\langle\psi_a(0)|U_{bc}(C)|\psi_c(0)\rangle\langle\psi_d(0)|\notag\\
&\qquad \times U^*_{bd}(C)\mathcal{U}(\tau)/D\big].
\end{align}
Applying the cyclic property of the trace and the completeness relation $\sum_{b}U_{bc}(C)U^*_{bd}(C)=\delta_{cd}$, we simplify this expression to
\begin{align}\label{Eq:sWZ3}
\theta_{\text{WZ}}(C)
=&\arg\text{Tr}_-\big[\frac{1}{D}P^2_-(0)|\mathcal{U}(\tau)\mathcal{U}^\dag(0)\big]\notag\\
=&\arg\text{Tr}_-\big[\frac{1}{D}P_-(0)U_-(C)\big],
\end{align}
where we have used the fact $P^2_-=P_-$ with $P_-(0)=\sum_a|\psi_a(0)\rangle\langle\psi_a(0)|$ being the initial ground-state projector. For the case $D=2$, this reduces exactly to Eq.~(\ref{thetaWZ}), thereby establishing the equivalence between the two definitions.

\section{Uhlmann-Berry correspondence for non-degenerate cases}\label{app_corr}
The following proof of the correspondence between the $T\rightarrow 0$ Uhlmann phase and the Berry phase in non-degenerate systems are based on the following two conditions. First, the adiabatic theorem requires a finite energy gap ($R>0$) throughout the evolution to (i) ensure confinement to the instantaneous ground-state manifold, (ii) suppress non-adiabatic transitions between energy levels, and (iii) guarantee finite-valued connection forms. Second, the system must remain non-degenerate, otherwise the appearance of degeneracy requires the replacement of the Berry connection by the WZ connection matrix, and the scalar WZ phase may be considered as a representative of the latter.

In the case $D_0=1$ without any degeneracy, Eq.~(\ref{AU}) reduces to
\begin{align}
&\lim_{T \to 0} \mathcal{A}_{\text{U}} = \mathcal{A}_{\infty} \notag \\
=& -\sum_i \dif |i(t)\rangle \langle i(t)| + \sum_i \langle i(t) | \dif | i(t) \rangle |i(t)\rangle \langle i(t)|
\end{align}
in the zero-temperature limit.
Here $|i(t)\rangle$ are the instantaneous energy eigenstates parameterized by $\mathbf{R}(t)$.
We introduce the unitary evolution operator of instantaneous eigenstates:
\begin{align}
\tilde{\mathcal{D}}(t) = \sum_k |k(t)\rangle \langle k(0)|,
\end{align}
which satisfies $\tilde{\mathcal{D}}^\dag\tilde{\mathcal{D}} = \tilde{\mathcal{D}}\tilde{\mathcal{D}}^\dag = 1_N$. Its differential is
\begin{align}
\dif \tilde{\mathcal{D}} \tilde{\mathcal{D}}^{-1} = \dif \tilde{\mathcal{D}} \tilde{\mathcal{D}}^{\dagger} = \sum_k \dif |k(t)\rangle \langle k(t)|.
\end{align}
We also define the diagonal Berry connection matrix as
\begin{align}
\hat{\mathcal{A}}_{\text{B}} = \sum_k \mathcal{A}_{\text{B}k} |k(t)\rangle \langle k(t)|, \quad \mathcal{A}_{\text{B}k} = \langle k(t) | \dif | k(t) \rangle.
\end{align}
Thus, $\mathcal{A}_{\infty}$ simplifies to
\begin{align}\label{Ainfty}
\mathcal{A}_{\infty} = \hat{\mathcal{A}}_{\text{B}} - \dif \tilde{\mathcal{D}}\tilde{\mathcal{D}}^{-1}.
\end{align}
In the zero-temperature limit, the Uhlmann phase is
\begin{align}
\lim_{T \to 0} \theta_{\text{U}} = \arg \left( \langle 0(0) | \mathcal{P} \exp \left( -\oint_C \mathcal{A}_{\infty} \right) | 0(0) \rangle \right)
\end{align}
since $\rho(0) \approx |0(0)\rangle \langle 0(0)|$ and the ground state dominates the trace.
We define the path-ordered evolution operator
\begin{align}
U(t) = \mathcal{P} \exp \left( -\int_0^t \mathcal{A}_{\infty}(s)  \dif s \right)
\end{align}
satisfying
\begin{align}
\frac{\dif}{\dif t} U(t) = - \mathcal{A}_{\infty}(t) U(t), \quad U(0) = I.
\end{align}
The Uhlmann phase is then
\begin{align}
\theta_{\text{U}} \approx \arg \left( \langle 0(0) | U(\tau) | 0(0) \rangle \right).
\end{align}

At zero temperature, we consider the state $|\phi(t)\rangle = U(t) | 0(0) \rangle$ with initial condition $|\phi(0)\rangle = |0(0)\rangle$. Under adiabatic evolution, we assume $|\phi(t)\rangle$ remains in the instantaneous ground state subspace:
\begin{align}
|\phi(t)\rangle = c(t) |0(t)\rangle.
\end{align}
The evolution equation is
\begin{align}
\frac{\dif}{\dif t} |\phi(t)\rangle = \dot{c}(t) |0(t)\rangle + c(t) |\dot{0}(t)\rangle = - \mathcal{A}_{\infty}(t) c(t) |0(t)\rangle.
\end{align}
Therefore, $\mathcal{A}_{\infty} |0(t)\rangle$ can be evaluated by using Eq.~(\ref{Ainfty}):
\begin{align}
\mathcal{A}_{\infty} |0(t)\rangle &= \left[ \sum_k \langle k | \dif k \rangle |k\rangle \langle k| - \sum_k \dif |k\rangle \langle k| \right] |0(t)\rangle \notag \\
&= \langle 0 | \dif 0 \rangle |0\rangle - \dif |0\rangle
\end{align}
since $\langle k|0\rangle = \delta_{k0}$. 
The evolution equation becomes
\begin{align}
\dot{c} |0\rangle + c |\dot{0}\rangle = c \left( - \langle 0 | \dot{0} \rangle |0\rangle + |\dot{0}\rangle \right).
\end{align}
Projecting the expression onto $|0(t)\rangle$ then leads to
\begin{align}
\dot{c} = - c \langle 0 | \dot{0} \rangle.
\end{align}
Meanwhile, projecting into the orthogonal components leads to trivial equalities.
Solving $\dot{c} = - c \langle 0 | \dot{0} \rangle$ with $c(0) = 1$ yields
$c(t) = \exp \left( -\int_0^t \langle 0(s) | \dif | 0(s) \rangle \right)$.
For a closed path $C$ with $\mathbf{R}(\tau) = \mathbf{R}(0)$, $|0(\tau)\rangle = |0(0)\rangle$ and
\begin{align}
c(\tau) = \exp \left( -\oint_C \langle 0 | \dif | 0 \rangle \right)= \exp \left( \mi \oint_C \mathcal{A}_{\text{B}0} \right) = \me^{\mi \theta_{\text{B}0}}.
\end{align}
Here $\mathcal{A}_{\text{B}0} = \mi \langle 0 | \dif | 0 \rangle$ is the Berry connection, and
$\theta_{\text{B}0} = \mathlarger{\oint}_C \mathcal{A}_{\text{B}0}$ is the Berry phase of the non-degenerate ground-state. Therefore,
$\langle 0(0) | \phi(\tau) \rangle = \langle 0(0) | c(\tau) | 0(0) \rangle = \me^{\mi \theta_{\text{B}0}}$,
and finally
\begin{align}
\lim_{T \to 0} \theta_{\text{U}} = \arg \left( \me^{\mi \theta_{\text{B}0}} \right) = \theta_{\text{B}0} \mod 2\pi.
\end{align}
This proof establishes the correspondence without requiring $[\hat{\mathcal{A}}_{\text{B}}, \dif \mathcal{D} \mathcal{D}^{-1}] = 0$, relying only on adiabatic evolution of the pure states and the zero-temperature limit of the Uhlmann phase.

\section{Proof of the zero-field Axis conditions}\label{app5}
Here we consider a quantum system with a degenerate ground-state subspace of dimension $D$ described by a principal $G$-bundle over the parameter space $M$, where $G = U(D)$. The WZ connection $\mathcal{A}_{\text{WZ}}$ is a $\mathfrak{g}$-valued 1-form ($\mathfrak{g} = \mathfrak{u}(D)$) satisfying the structure equation $\mathcal{F}_{\text{WZ}} = \dif\mathcal{A}_{\text{WZ}} + \mathcal{A}_{\text{WZ}} \wedge \mathcal{A}_{\text{WZ}}$,
where $\mathcal{F}_{\text{WZ}}$ is the curvature 2-form. The connection induces parallel transport along paths in $M$. For simplicity, we will omit the subscript ``WZ'' in the following discussion.

\subsection{Proof of the vanishing connection}
Let $\mathcal{S} \subset M$ be a submanifold satisfying the following conditions. \textbf{Symmetry:} $\mathcal{S}$ is invariant under a connected Lie subgroup $\mathcal{H} \subset G$ with Lie algebra $\mathfrak{h}$. \textbf{Cartan Alignment:} $\mathcal{S}$ is an integral manifold of the Cartan subalgebra $\mathfrak{t} \subset \mathfrak{g}$, where $\mathfrak{t}$ is the maximal Abelian subalgebra.

Since $\mathcal{S}$ is an integral manifold of $\mathfrak{t}$, the restriction $\mathcal{A}|_{\mathcal{S}}$ takes values in $\mathfrak{t}$ as
\begin{align}
\mathcal{A}|_{\mathcal{S}} \in \Omega^1(\mathcal{S}, \mathfrak{t}).
\end{align}
The Cartan subalgebra $\mathfrak{t}$ is Abelian, so the curvature simplifies to
\begin{align}
\mathcal{F}|_{\mathcal{S}} = \dif\mathcal{A}|_{\mathcal{S}}.
\end{align}
The symmetry condition implies that $\mathcal{H}$ preserves the physical state along $\mathcal{S}$. As $\mathcal{H}$ acts trivially on $\mathcal{S}$ (by invariance), the holonomy along any contractible loop in $\mathcal{S}$ must commute with $\mathcal{H}$. Since $\mathcal{H}$ is connected and $\mathfrak{t}$ is maximal Abelian, the holonomy lies in $\exp(\mathfrak{t})$.

Next, we consider a local orthonormal frame $\{e_j\}$ for the ground-state bundle over $\mathcal{S}$. The connection coefficients are
\begin{align}
\mathcal{A}_{jk} = \langle e_j | \dif e_k \rangle.
\end{align}
The symmetry condition requires that $\mathcal{H}$ preserves the frame
\begin{align}
h \cdot e_j = \sum_k U_{jk}(h) e_k, \quad \forall h \in \mathcal{H},
\end{align}
where $U: \mathcal{H} \to U(D)$ is a unitary representation. Differentiation at the identity gives
\begin{align}
\xi \cdot e_j = \sum_k \Lambda_{jk}(\xi) e_k, \quad \xi \in \mathfrak{h},
\end{align}
where $\Lambda: \mathfrak{h} \to \mathfrak{u}(D)$ is the derived representation. The infinitesimal parallel-transport condition is
\begin{align}
\langle e_j | \xi \cdot e_k \rangle + \langle \xi \cdot e_j | e_k \rangle = 0.
\end{align}
Plugging the derived representation into it, one obtains
\begin{align}
\Lambda_{kj}(\xi) + \Lambda^*_{jk}(\xi)= 0,
\end{align}
which holds automatically since $\Lambda(\xi) \in \mathfrak{u}(D)$. The connection form must satisfy
\begin{align}
\mathcal{L}_\xi \mathcal{A} = 0, \quad \forall \xi \in \mathfrak{h},
\end{align}
where $\mathcal{L}_\xi$ is the Lie derivative. In components,
\begin{align}
(\mathcal{L}_\xi \mathcal{A})_{jk} = \xi \cdot \mathcal{A}_{jk} - \mathcal{A}([\xi, \cdot])_{jk}.
\end{align}
As $\mathcal{S}$ is $\mathcal{H}$-invariant and $\mathcal{A}|_{\mathcal{S}} \in \mathfrak{t}$, the $\mathfrak{t}$-valuedness and $\mathcal{H}$-invariance imply $\mathcal{A}|_{\mathcal{S}}$ is constant. Since $\mathcal{S}$ is connected, a gauge transformation can rotate $\mathcal{A}|_{\mathcal{S}}$ to zero. Specifically, one solves the differential equation
\begin{align}
\dif g = -g \mathcal{A}|_{\mathcal{S}}, \quad g(p_0) = I,
\end{align}
along paths in $\mathcal{S}$. As $\mathcal{A}|_{\mathcal{S}}$ is $\mathfrak{t}$-valued and Abelian, the solution is
\begin{align}
g(p) = \exp\left(-\int_{p_0}^p \mathcal{A}|_{\mathcal{S}}\right).
\end{align}
Therefore, the transformed connection is
\begin{align}
\mathcal{A}' = g^{-1}\dif g + g^{-1}\mathcal{A}g = 0.
\end{align}
Consequently, $\mathcal{A}_{\text{WZ}}|_{\mathcal{S}} = 0$ in this gauge.

\subsection{Application to the 4D tight-binding model}
In the 4D tight-binding model, the path
$\mathcal{S}:\, k_y=k_z=k_u=0$
is the fixed-point set of the residual $U(1)\cong SO(2)$ rotation acting on the $(k_y,k_z,k_u)$ subspace, generated by
$J_{\mathbf n}=n_yJ_y+n_zJ_z+n_uJ_u$ with
$J_y=k_u\partial_{k_z}-k_z\partial_{k_u}$ and its cyclic permutations.
Along $\mathcal{S}$, the Hamiltonian reduces to
$H_{\mathcal{S}}(k_x)=\sin k_x\Gamma^{1}+(m+3+\cos k_x)\Gamma^{5}$
and commutes with every $J_{\mathbf n}$ on $\mathcal{S}$:
\begin{align}
\bigl[H_{\mathcal{S}},J_{\mathbf n}\bigr]\big|_{\mathcal{S}}=0
\end{align}
since $J_{\mathbf{n}}$ acts exclusively on $(k_y, k_z, k_u)$ while $H_{\mathcal{S}}$ depends solely on $k_x$.
Consequently, the symmetry acts according to
$R(\theta)H_{\mathcal{S}}R^{-1}(\theta)=H_{\mathcal{S}}$ with
$R(\theta)=\exp(\theta J_{\mathbf n})$.

Because the symmetry acts trivially on the ground states along $\mathcal{S}$, the bundle splits into two one-dimensional sub-bundles. Each has a structural group of the effective one-dimensional Cartan sub-algebra
$\mathfrak{t}_{\text{eff}}=\operatorname{span}\{\mi\sigma_z\}\subset\mathfrak{u}(2)$.
Hence $\mathcal{S}$ is an integral curve of $\mathfrak{t}_{\text{eff}}$, and the curvature reduces to the Abelian form
$\mathcal{F}|_{\mathcal{S}}=\mathrm d\mathcal{A}|_{\mathcal{S}}$.
Thus, a global gauge rotation
$g(k_x)=\exp\Bigl(-\int_{0}^{k_x}\mathcal{A}|_{\mathcal{S}}\Bigr)$
sets the WZ connection to zero:
$\mathcal{A}_{\text{WZ}}\big|_{\mathcal{S}}=0$.
The vanishing connection is a direct consequence of the residual $U(1)$ symmetry and the resulting Abelian structure of the effective Cartan sub-algebra. This gives rise to the zero-field axis.

\section{Evaluations of $\theta_\text{U}$ under different orders of taking limits}\label{app4}
To derive the result $\lim_{m \to -4^+} \left( \lim_{T \to 0} \theta_{\text{U}}(m) \right) = \pi$ for the 4D tight-binding model, we analyze the Uhlmann phase along the path $C(t) = (k_x(t), 0, 0, 0)$ given by $\theta_{\text{U}} = \arg \left( \cos I(m, T) \right)$ with $I(m, T) = \mathlarger{\int}_0^{2\pi} f(k_x, T)  \dif k_x$. The kernel function $f$ is given by Eq.~\eqref{kernelf}.
First, we take the zero-temperature limit ($T \to 0$) for fixed $m > -4$. In this case, $\beta \to \infty$ and $\operatorname{sech}(\beta R) \to 0$, simplifying the integral to
\begin{align}
\lim_{T \to 0} I(m, T)=I_0(m)= -\frac{1}{2} \int_0^{2\pi} \frac{(m + 3) \cos k_x + 1}{R^2}  \dif k_x.
\end{align}
Introducing $a = m + 3$ and $z = \me^{\mi k_x}$, the expression becomes
\begin{align}
I_0(m) &= -\frac{1}{4i} \oint_{|z|=1} \frac{a(z^2 + 1) + 2z}{z (a z^2 + (a^2 + 1) z + a)}  \dif z.
\end{align}
The poles are located at $z = 0$, $z = -a$, and $z = -1/a$. For $m > -4$ ($a > -1$) and $m < -2$ ($a < 1$), the poles inside $|z| = 1$ are $z = 0$ and $z = -a$, whose residues are evaluated as follows:
\begin{align}\label{z0}
\operatorname{Res} (z=0)= \lim_{z \to 0} z \cdot \frac{a(z^2 + 1) + 2z}{z (a z^2 + (a^2 + 1) z + a)} = 1,
\end{align}
and
\begin{align}
\operatorname{Res}(z=-a) = \lim_{z \to -a} (z + a) \cdot \frac{a(z^2 + 1) + 2z}{z (z + a)(a z + 1)} = 1.
\end{align}
Therefore, $I_0(m) = -\frac{1}{4\mi} \cdot 2\pi \mi(1+1) = -\pi$, and the zero-temperature limit of the Uhlmann phase is
\begin{align}
\lim_{T \to 0} \theta_{\text{U}}= \arg \left( \cos (-\pi) \right) = \pi.
\end{align}
Consequently, $\lim_{m \to -4^+} \left( \lim_{T \to 0} \theta_{\text{U}}(m) \right) =  \pi$.

If $m<-4$ or $m\to -4^-$, then $|a| < 1$. The poles inside $|z|=1$ are $z_0 = 0$ and $z_2 = -1/a$ ($|z_2| = 1/|a| < 1$). The residue at $z=0$ has been evaluated in Eq.~(\ref{z0}). The residue at $z_2 = -1/a$ is given by
\begin{align}
\lim_{z\to -1/a} (z + 1/a) \cdot \frac{a(z^2+1) + 2z}{z (z + a)(z + 1/a)}=  -1.
\end{align}
The sum of residues is $1 + (-1) = 0$. Thus, $I_0(m) =0$, which yields $\lim_{m \to -4^-} \left( \lim_{T \to 0} \theta_{\text{U}}(m) \right) =  \arg\left( \cos (0) \right)=0 $. The analysis explains the jumps of the $T\rightarrow 0$ Uhlmann phase across the Dirac points which do not appear in the scalar WZ phase.

%

\end{document}